\documentclass[12pt, preprint]{article}
\usepackage{jcappub}
\usepackage{amsmath}
\usepackage{graphicx}
\usepackage{multicol}
\usepackage{color}
\usepackage{subfig}
\usepackage{caption}
\usepackage{hyperref}
\usepackage{mathtools}
\usepackage{tablefootnote}
\usepackage{soul}
\usepackage{ulem}
\usepackage{float}
\usepackage{multirow}
\usepackage{longtable}
\usepackage[usenames,dvipsnames,svgnames,table]{xcolor}

\title{Growth of Massive Black Hole Seeds by Migration of Stellar and Primordial Black Holes: Gravitational Waves and Stochastic Background}

\author[a,b,c]{Lumen Boco,}
\author[a,b,c,d]{Andrea Lapi,}
\author[a]{Alex Sicilia,}
\author[a,b,c]{Giulia Capurri,}
\author[a,b,c]{Carlo Baccigalupi,}
\author[a,c]{Luigi Danese}

\affiliation[a]{SISSA, via Bonomea 265, 34136, Trieste, Italy}
\affiliation[b]{INFN, Sezione di Trieste, via Bonomea 265, 34136, Trieste, Italy}
\affiliation[c]{IFPU, via Beirut 2, 34151, Trieste, Italy}
\affiliation[d]{INAF/IRA, via Piero Gobetti 101, 40129, Bologna, Italy}

\emailAdd{lboco@sissa.it}
\emailAdd{lapi@sissa.it}
\emailAdd{asicilia@sissa.it}
\emailAdd{gcapurri@sissa.it}
\emailAdd{bacci@sissa.it}
\emailAdd{danese@sissa.it}

\abstract{We investigate the formation and growth of massive black hole (BH) seeds in dusty star-forming galaxies, relying and extending the framework proposed by \citep{boco+20}. Specifically, the latter envisages the migration of stellar compact remnants (neutron stars and stellar-mass black holes) via gaseous dynamical friction towards the galaxy nuclear region, and their subsequent merging to grow a massive central BH seed. In this paper we add two relevant ingredients: (i) we include primordial BHs, that could constitute a fraction $f_{\rm pBH}$ of the dark matter, as an additional component participating in the seed growth; (ii) we predict the stochastic gravitational wave background originated during the seed growth, both from stellar compact remnant and from primordial BH mergers. We find that the latter events contribute most to the initial growth of the central seed during a timescale of $10^6-10^7\,\rm yr$, before stellar compact remnant mergers and gas accretion take over. In addition, if the fraction of primordial BHs $f_{\rm pBH}$ is large enough, gravitational waves emitted by their mergers in the nuclear galactic regions could be detected by future interferometers like Einsten Telescope, DECIGO and LISA. As for the associated stochastic gravitational wave background, we predict that it extends over the wide frequency band $10^{-6}\lesssim f [{\rm Hz}]\lesssim 10$, which is very different from the typical range originated by mergers of isolated binary compact objects. On the one hand, the detection of such a background could be a smoking gun to test the proposed seed growth mechanism; on the other hand, it constitutes a relevant contaminant from astrophysical sources to be characterized and subtracted, in the challenging search for a primordial background of cosmological origin.}


\begin{document}
\maketitle
\flushbottom

\section{Introduction}
The discovery of gravitational waves (GWs) by the LIGO/Virgo team \citep{abbott+16}, and the subsequent detection of many other GW signals \citep{abbott+19a, abbott+20b} has largely impacted on different research fields: on astrophysics, with the discovery of a new class of massive black holes (BHs) \citep{abbott+19b, abbott+20a, abbott+20c}, and with a tentative first characterization of the BH mass function and of the double compact objects merging rates \citep{abbott+20c}; on cosmology, with the first independent estimation of the Hubble constant \citep{abbott+17a, abbott+21, hotokezaka+19, soares-santos+19, mukherjee+21}; on fundamental physics, with the measurement of the GW propagation speed that has ruled out some modified gravity theories \citep{abbott+17b, creminelli+17}. This is only the tip of the iceberg, since GW astronomy will become increasingly precise and robust, with the advent of the next observing runs for Advanced LIGO/VIRGO, with KAGRA entering in the network of ground based interferometers, and with the future third generation detectors such as the Einstein Telescope (ET), the Deci-hertz Interferometer Gravitational wave Observatory (DECIGO), and the Laser Interferometer Space Antenna (LISA), which will provide an increased redshift depth and new frequency observational windows. 

Despite such a head start, GW astronomy has still to address some major issues. One concerns the main astrophysical channels that can lead to the formation and merging of double compact objects. Although many works have been focused on the merging rates and mass spectrum of isolated compact objects binaries (e.g., \citep{belczynski+16, lamberts+16, cao+18, elbert+18, li+18, mapelli+18, artale+19, boco+19, neijssel+19, boco+21, santoliquido+21}), the rates of compact objects mergers due to dynamical interactions in dense environments such as globular or nuclear star clusters are still largely unknown. In addition, the possible existence of primordial black holes (pBHs) and pBH binaries may constitute additional sources of GW emission and GW background production (e.g. \citep{clesse+17, raidal+17, scelfo+18, kimura+21, wang+16, wang+19}). In the next decades, with new runs of current GW interferometers and with the advent of future detectors, there should be the possibility to better characterize the GW signals and to acquire enough statistics to test, and eventually confirm or rule out some astrophysical models of GW production.

In this vein, \citep{boco+20} have proposed a mechanism for the growth of massive BH seeds in the central regions of dusty star-forming galaxies; these are thought to be the progenitors of local massive spheroidal galaxies, that host relic supermassive BHs at their centers. In short, the framework envisages the migration of stellar compact remnants (e.g., neutron stars and BHs) towards the nuclear galaxy regions via dynamical friction against the dense gaseous environment, and their subsequent mergers to grow a massive central BH seed. The process is particularly efficient in dusty star-forming galaxies because they feature large star formation rates (SFR) $\psi\gtrsim 100-1000\,\rm M_\odot/yr$ and huge molecular gas reservoirs $\rm M_{\rm gas}\gtrsim 10^{10}-10^{11}\,M_\odot$ concentrated in a compact region of a few $\rm kpc$ (e.g., \citep{scoville+14, scoville+16, ikarashi+15, simpson+15, barro+16, spilker+16, tadaki+17a, tadaki+17b, tadaki+18, talia+18, lang+19, smail+21}). 
These conditions are prompt for the efficient sinking of innumerable compact objects toward the nuclear regions via gaseous dynamical friction. In fact, \citep{boco+20} have demonstrated that this mechanism is able to grow heavy BH seeds of masses $\sim 10^4-10^6\,M_\odot$ within some $10^7\,\rm yr$, so possibly alleviating the problem of supermassive BH formation at high redshift. Even more, \citep{boco+20} have proposed a clear-cut way of testing their framework: the continuous mergers of migrating stellar compact remnants with the growing central BH seed will produce GW signals with precise signatures, that could be within the reach of future interferometers such as ET, DECIGO and LISA. 

In the present work we extend the analysis and the predictions by \citep{boco+20} under two respects:
\begin{itemize}
    \item We add pBHs into the game. If pBHs are present and constitute a fraction $f_{\rm pBH}$ of the dark matter (DM) mass, they will undergo the gaseous dynamical friction process, especially in the central galactic region where the gas is more concentrated. Therefore they could contribute to the growth of the central BH seed; moreover, their mergers with said seed could produce GW signals with specific properties. We study these two effects as a function of the pBH fraction $f_{\rm pBH}$.

    \item We make predictions regarding the stochastic gravitational wave background (SGWB) produced by all the unresolved merger events, both for stellar compact remnants and for pBHs. We will show that the detection of the SGWB over an extended frequency spectrum could constitute a crucial test for our scenario of seed formation.
\end{itemize}

The paper is organized as follows: in section \ref{sec:DF} we briefly recap the mechanism of gaseous dynamical friction, with particular focus on the estimate of the related timescale; we also describe the physical setup, and we compute the merging rates of stellar compact remnants and pBHs; in section \ref{sec:seed} we discuss the ensuing growth of the central BH mass as a function of time; in section \ref{sec:gw} we compute the GW emission rates and discuss their detectability with ET, DECIGO and LISA; in section \ref{sec:sgwb} we predict the SGWB originated by the incoherent superposition of unresolved merging events. Finally, in section \ref{sec:conclusion} we summarize our findings and outlook future developments.

Throughout this work, we adopt the standard flat $\Lambda$CDM cosmology (see \citep{planck2020}) with rounded parameter values: matter density $\Omega_M=0.32$, baryon density $\Omega_b=0.05$ and Hubble constant $H_0=100\, h$ km s$^{-1}$ Mpc$^{-1}$ with $h = 0.67$.

\section{Gaseous dynamical friction migration timescales and merging rates}\label{sec:DF}

Dynamical friction is the process of interaction between an object (the perturber) moving in a sea of background particles, and its gravitationally induced wake. The ensuing energy and angular momentum loss generally cause an orbital decay of the perturber. Dynamical friction against a background of collisionless particles such as stars or dark matter has been vastly discussed in literature in many contexts (e.g., \citep{chandrasekhar+43, binney+87, lacey+93, hashimoto+03, fujii+06, boylan-kolchin+08, jiang+08}), as e.g. in the formation of (super)massive BH binaries after a galaxy merger (see \citep{begelman+80, mayer+07, barausse+12, chapon+13, antonini+15, tamburello+17, dayal+19, katz+20}). On the other hand, dynamical friction against a background of collisional particles, such as a gaseous environment, has drawn much less attention. Yet, a series of classical works (e.g., \citep{dokuchaev+64, ruderman+17, bisnovatyi-kogan+79, rephaeli+80, ostriker+99}) found that, for a perturber in supersonic motion, gaseous dynamical friction can be as efficient as that occurring against a collisionless background. 
The dynamical friction force $F_{\rm DF}$ can be analytically expressed as:
\begin{equation}
F_{\rm DF}=-\frac{4\pi\,G^2\,m_\bullet^2\,\rho_{\rm gas}}{v^2}f(\mathcal{M}, \ln\Lambda)
\end{equation}
where $\rho_{\rm gas}$ is the background gas density, $m_\bullet$ is the mass of the perturber, $v$ its velocity and $f (\mathcal{M}, \ln\Lambda)$ is a function of the Mach number $\mathcal{M}\equiv v/c_s$, i.e. the ratio of the perturber velocity to the sound speed $c_s$ of the background medium, and of the Coulomb logarithm $\ln\Lambda$. In the collisional case, the reference formula for $f(\mathcal{M},\ln\Lambda)$ has been derived in \citep{ostriker+99}, studying a perturber on a straight motion in a uniform gas distribution via time-dependent linear perturbation theory. The main result is that for bodies moving at supersonic speeds, the gaseous drag force is more efficient than in the case of collisionless medium and that even for subsonic motion the gravitational drag does not completely vanishes. Numerical confirmations of these results have been provided by \citep{sanchez-salcedo+01, escala+04}, extending the results of \citep{ostriker+99} also to perturbers on non-straight trajectories. They studied the orbital decay of a moving object in a uniform gaseous medium and found a pleasant agreement with the analytical formula of \citep{ostriker+99}, apart for an overestimation of the decay timescale of a factor $\sim 1.5$ for perturbers in slightly supersonic motion $\mathcal{M}\sim \sqrt{2}\,c_s$. In \citep{escala+04} it is proposed a new parametric formula to correct for this overestimation. Subsequent works (\citep{tanaka+09, tagawa+16}) refined the computation providing an expression for $f(\mathcal{M},\ln\Lambda)$ that we took as a reference in \citep{boco+20} and in the present work. Even though the general shape of the dynamical friction force and the expression for $f(\mathcal{M},\ln\Lambda)$, besides small discrepancies, appear to be in rather good agreement among different authors, the value of the Coulomb logarithm $\ln\Lambda$ is still somewhat debated. This is extensively discussed in \citep{boco+20} (see references therein) where the authors tried different prescriptions for the Coulomb logarithm, checking that the results do not strongly depend on it. We adopt here the reference model of \citep{boco+20}, but we refer the interested reader to the analysis reported there to check the impact of different choices.

To put the mechanism in astrophysical context, our main aim is to describe the formation of the seeds for (super)massive BHs. In the local Universe these giant monsters are hosted at the center of massive spheroidal galaxies. Thus their seeds must have formed in the progenitors of such systems at intermediate/high-redshift, which are known to be dusty star-forming galaxies.
These objects, detected and investigated mainly in the far-IR/(sub)mm band by ground-based interferometers like ALMA, feature large SFRs $\psi\gtrsim 100-1000\,\rm M_\odot/yr$ and huge molecular gas reservoirs $\rm M_{\rm gas}\gtrsim 10^{10}-10^{11}\,M_\odot$ concentrated in a compact region of a few $\rm kpc$ (e.g., \citep{scoville+14, scoville+16, ikarashi+15, simpson+15, barro+16, spilker+16, tadaki+17a, tadaki+17b, tadaki+18, talia+18, lang+19, smail+21}). 
These conditions are prompt for the efficient sinking of many compact objects toward the nuclear regions via the aforementioned process of gaseous dynamical friction. In \citep{boco+20} we have run a series of dynamical simulations and derived a fitting formula for the corresponding migration timescale:
\begin{equation}
\tau_{\rm DF}=\mathcal{N}\left(\frac{m_\bullet}{100\,\rm M_\odot}\right)^a \left(\frac{M_{\rm gas}}{10^{11}\,\rm M_\odot}\right)^b \left(\frac{R_{\rm e}}{1\,\rm kpc}\right)^c \left(\frac{j}{j_c(\epsilon)}\right)^\beta \left(\frac{r_c(\epsilon)}{10\,\rm pc}\right)^\gamma
\label{eq:tauDF}
\end{equation}
The quantities entering the above expression are related to the properties of the compact objects and of the gas distribution in the galaxy: $M_{\rm gas}$ is the total gas mass; $R_e$ is the half mass radius of the gas distribution; $m_\bullet$ is the mass of the migrating compact object; $\epsilon$ and $j$ are the initial specific energy and angular momentum of the compact object; $r_c(\epsilon)$ is the circular radius that the compact object would have if it were on a circular orbit with energy $\epsilon$, and $j_c(\epsilon)$ is the angular momentum associated to that orbit, measuring the circularity of the orbit (typical values for circular radius and circularity can be found in Table 2 of \citep{boco+20}). The precise values of the exponents $(a,b,c,\beta,\gamma)$ and of the normalization factor $\mathcal{N}$ depend on the specific shape of the gas density profile (see Table 1 in \citep{boco+20}). In the present work we adopt the fiducial setup of \citep{boco+20}, namely, a 3D Sersic gas density profile $\rho(r)\propto r^{-\alpha}\, e^{-k\,(r/R_e)^{1/n}}$ with $n=1.5$, $\alpha=1-1.188/2n+0.22/4n^2\sim 0.6$ (see \citep{prugniel+97}) and half-mass radius $R_{\rm e}\sim 1$ kpc. Then the values for the parameters in equation (\ref{eq:tauDF}) read $a\approx-0.95$, $b\approx0.45$, $c\approx-1.2$, $\beta\approx 1.5$, $\gamma\approx 2.5$ and $\mathcal{N}\approx 3.4\times 10^8$ yr. 

We caveat that here we are assuming a smooth gaseous distribution which can be easily modeled analytically. However, in principle, the gas could present a clumpy structure with overdense and underdense regions randomly distributed. This would surely modify the dynamical friction timescale, with dense gaseous clumps exerting a stronger drag with respect to underdense regions. While the exact effect on the perturber motion of this complex gaseous structure can be examined only via hydrodynamical simulations, a discussion about this issue is presented in \citep{boco+20} (see Section 5 there); we report here some highlights. First of all, as shown in \citep{boco+20}, the considered process acts on small scales $\lesssim 300\,\rm pc$; therefore large scales clumpiness of the gaseous medium should not impact on it. On the other hand, star-forming molecular clouds with radii $\sim 10-20\,\rm pc$ could be present in the central region. However, observations show a rather smooth distribution of the stellar mass in high-z star-forming systems (e.g., Swinbank et al. 2010; Hodge et al. 2016; Rujopakarn et al. 2016; Lang et al. 2019) and in their quiescent descendants (e.g., van der Wel \& van der Marel 2008; Belli et al. 2017; Cappellari et al. 2013), indicating that molecular clouds are dissolved or a substantial amount of stars can escape quite rapidly from them. Still, some compact remnants born within the cloud might remain bound to a stellar cluster that originated there, reducing the amount of remnants available for the central BH growth. However, during the formation of the bulge, the stellar clusters may themselves migrate toward the central region via dynamical friction against the background stars and contribute to the growth of a nuclear star cluster there (e.g., Antonini et al. 2015). Given the short lifetime/high escape fraction of these small-scale clumpy structures, we expect them to impact only marginally on the whole mechanism efficiency.

Next we describe how the dynamical friction timescale is employed to derive the merging rates of compact remnants from stellar evolution and of pBHs.

\subsection{Merging rates of stellar compact remnants}\label{subsec:stellar_param}

Given the high SFR ongoing in the progenitors of local spheroidal galaxies, a lot of stars and compact remnants are formed in a short timescale within the nuclear regions. We assume that stars are initially distributed in space as the gas density profile $\rho$; thus the probability distribution for a star to be born at distance $r$ from the galactic center is
\begin{equation}
\frac{{\rm d}p}{{\rm d}r}\propto r^2\, \rho(r)\;.
\label{eq:positions}
\end{equation}
After $\lesssim 10^7\,\rm yr$ massive stars ($m_\star\gtrsim 7-8\,M_\odot$) undergo a supernova explosion leaving a compact remnant, such as a neutron star or a stellar-mass BH. We assume that compact remnants follows the same velocity distribution of the progenitor stars, which is in turn related to that of the star-forming molecular gas cloud. We assume a Gaussian distributions of radial and tangential velocities:
\begin{equation}
\frac{{\rm d}p}{{\rm d}v_{r,\theta}}\propto e^{-v_{r,\theta}^2/2\sigma^2}\;,
\label{eq:velocities}
\end{equation}
with dispersion $\sigma(r)$ found by solving the isotropic Jeans equation:
\begin{equation}
    \sigma^2(r)\propto \frac{1}{\rho(r)}\int_r^\infty\,{\rm d}r'\frac{\rho(r')}{r'^2}\, \int_0^{r'}{\rm d}r''\, r''^2\,\rho(r'')\;,
    \label{eq:sigma}
\end{equation}
attaining values $\sigma(r)\simeq 150-300\,\rm km\,s^{-1}$ for initial radii $r\simeq 10-100\,\rm pc$. From the distributions in equations (\ref{eq:positions}) and (\ref{eq:velocities}) the initial positions and velocities of stellar compact remnants, their initial energy and angular momentum can be easily extracted.

It will be convenient to characterize a galaxy by its spatially and temporally averaged SFR $\psi$; this is because in the sequel, when computing cosmic merger rates and associated gravitational wave emission, we will exploit the galaxy statistics based on this quantity (see section \ref{sec:gw}). Thus, for a given SFR $\psi$ we first compute the associated stellar mass $M_\star$ from the well-established galaxy main sequence relationships (see \citep{speagle+14}) and then estimate the initial gas mass $M_{\rm gas}$, entering the dynamical friction timescale, from the redshift-dependent $M_{\rm gas}-M_\star$ relation by \citep{lapi+17a} (see also \citep{moster+13, aversa+15, shi+17, behroozi+19}) based on abundance matching techniques.

The merging rates per unit remnant mass in a galaxy with spatially integrated SFR $\psi$ at redshift $z$ can be computed as (see \citep{boco+20} for details):
\begin{equation}
\frac{{\rm d}\dot{N}_{{\rm DF},\star}}{{\rm d}m_\bullet}(m_\bullet,\tau|\psi,z)=\int{{\rm d}r}\,\frac{{\rm d}p}{{\rm d}r}(r)\,\int{{\rm d}v_\theta}\,\frac{{\rm d}p}{{\rm d}v_\theta}(v_\theta|r)\,\int{{\rm d}v_r}\,\frac{{\rm d} p}{{\rm d} v_r}(v_r|r)\, \frac{{\rm d}\dot{N}_{\rm birth}}{{\rm d}m_\bullet}(m_\bullet, \tau-\tau_{\rm DF}|\psi, z)
\label{eq:merging_stellar}
\end{equation}
where ${\rm d}\dot{N}_{\rm birth}/{\rm d}m_\bullet(m_\bullet,\tau|\psi,z)$ is the birth rate of a compact remnant of mass $m_\bullet$ at the galactic age $\tau$, computed as in \citep{boco+19} combining prescriptions of galactic and stellar evolution; $\tau_{\rm DF}$ is the dynamical friction timescale after equation (\ref{eq:tauDF}). The rationale behind this expression is that the merging rates of migrating compact remnants that contribute to the growth of the central BH seed at a time $\tau$ depends on the birthrates of such objects at a time $\tau-\tau_{\rm DF}$, weighted by the corresponding distributions of initial positions and velocities.

We close the section highlighting some possible caveats related to supernova (SN) explosions which might modify the dynamical friction timescale and the ensuing merging rates. Stellar compact remnants, indeed, are originated after SN explosions which could remove a sizeable amount of gas from their surroundings hampering the dynamical friction process. In addition, SN explosions could be asymmetric, expelling more material in a certain direction with respect to others, and consequently imprinting a momentum kick to the compact remnant, changing its velocity with respect to the progenitor star and breaking our hypothesis on the initial velocity distribution.

However, both these effects are strongly mitigated in our context by the fact that compact remnants which mainly contribute to the merging rates and to the growth of the central BH, especially in the initial phase, feature huge masses $m_\bullet\gtrsim 30\,\rm M_\odot$ (see Figure 3 in \citep{boco+20}). These massive BHs are produced at low metallicities by stars with $m_\star>30-35\,\rm M_\odot$ which are characterized by large fallback fractions $f_{\rm fb}$; in other words, a large fraction of the envelope mass falls back onto the core and contributes to the BH formation, so reducing the power of SN feedback and the natal kick momentum \citep{belczynski+08, dominik+12}. In particular, in \citep{belczynski+10} it is estimated that stars with $m_\star\gtrsim 40\,\rm M_\odot$ have $f_{\rm fb}\sim 1$, undergoing a direct collapse characterized by no explosion and zero natal kick. Therefore the main contributors to the merging rates and to the central BH growth should be scarcely affected by these effects.

Still, since these processes may affect the evolution of lower mass BHs and neutron stars (NS) at later times, an order-of-magnitude estimate of their impact is in order. As for the SN explosion, it can efficiently sweep up material
during the energy-conserving expansion phase, out
to a radius $R_{\rm SN}\sim 5\, t_4^{2/(5-\alpha)}\,n_{2}^{-1/(5-\alpha)}\, E_{51}^{1/(5-\alpha)}\, \rm pc$ where $E_{51}\equiv E_{\rm SN}/10^{51}\, \rm erg$ is the explosion energy, $n_2\equiv n/10^2\,\rm cm^{-3}$ is the average gas density and $t_4\equiv t/10^4\,\rm yr$ the time since the explosion (e.g., Ostriker \& McKee 1988; Mo et al. 2010); however, once formed the remnant will move in the gaseous medium at a typical velocity of $\sigma_{200}\equiv \sigma/200\,\rm km\, s^{-1}$ and thus will travel a distance $R_{\rm rem}\sim 2\, \sigma_{200}\,t_4\,\rm pc$, implying that most of the gas mass swept up by the remnant is replaced after $\lesssim 10^5$ yr. The natal kick, instead, can be estimated (following \citep{mapelli+21}) as: $v_{\rm kick}\simeq v_H\,\langle m_{\rm NS}\rangle/m_\bullet$, where $v_H$ is drawn from a Maxwelian distribution with $\sigma_H\sim 265\,\rm km\,s^{-1}$, observationally derived from the motion of pulsars in the Galaxy (\citep{hobbs+05}), and $\langle m_{\rm NS}\rangle=1.33$ is the average NS mass. The resulting kick velocity is of the order of $\sim 30\,\rm km\,s^{-1}$ even for low mass BHs $m_\bullet\sim 10\,\rm M_\odot$, which is a factor $\sim 5-10$ below the typical velocities of the compact remnants considered $v\sim 150-300\,\rm km\,s^{-1}$. Still, for lower mass BHs and especially for NS the kick could have some impact on the dynamics, shortening or extending the dynamical friction timescale depending on the kick direction with respect to the initial velocity.

\subsection{Merging rates of primordial black holes}\label{subsec:primordial_param}

Primordial black holes (pBHs), if they exist, would also undergo the dynamical friction process and sink towards the nuclear galactic region, contributing to the growth of the central BH seed. We quantify the total number of pBHs $N_{\rm pBH}(\psi)$ present in a galaxy with average SFR $\psi$ by the expression
\begin{equation}
N_{\rm pBH}\approx \cfrac{f_{\rm pBH}\,M_{\rm H}}{\int{{\rm d} m_\bullet}\,m_\bullet\,\cfrac{{\rm d}p}{{\rm d}m_\bullet}}\;.
\end{equation}
In the above $M_{\rm H}(\psi)$ is the dark matter (halo) mass of the galaxy, $f_{\rm pBH}$ is the fraction of halo mass $M_{\rm H}$ constituted by pBHs, and ${\rm d}p/{\rm d}m_{\bullet}$ is the pBH mass distribution in terms of the individual pBH mass $m_\bullet$. We estimate the halo mass from the $M_{\rm H}-\psi$ relationship derived via abundance matching techniques (see \citep{aversa+15, scelfo+20, capurri+21}). The pBH mass functions is theoretically determined by the pBHs formation mechanism (e.g. \citep{sasaki+18, carr+10, carr+17, carr+21}), but largely unconstrained by (even indirect) observations. Stringent upper limits have been placed on $f_{\rm pBH}$, yet in turn still somewhat dependent on the mass function. In the present work we adopt a log-normal distribution of pBH masses with central value of $30\,M_\odot$ and dispersion $\sigma_{\log m_\bullet}=0.3$ dex. These have been selected to fall in a region of the parameter space where $f_{\rm pBH}$ is still poorly constrained  (see \citep{carr+17}), so as to allow maximal flexibility. 

The initial spatial distribution of pBHs is assumed to follow the DM density profile, in terms of a Navarro-Frank-White (1996; NFW) distribution $\rho_{\rm H}(r)\propto 1/r\,(r+r_s)^2$. As in the compact remnant case, the probability for a pBH to be born at a distance $r$ from the center is $\propto r^2\, \rho_{\rm H}(r)$ and we assume the radial and tangential velocities distributions to be Gaussians with dispersion computed as in equation (\ref{eq:sigma}) keeping into account the total density profile. The merging rates per unit pBH mass in a galaxy with average SFR $\psi$ at redshift $z$ is written \begin{equation}
\frac{{\rm d}\dot{N}_{\rm DF,pBH}}{{\rm d}m_\bullet}(m_\bullet,\tau|\psi,z)=N_{\rm pBH}\,\frac{{\rm  d}p}{{\rm d}m_\bullet}\,\int{{\rm d}r}\,\frac{{\rm d}p}{{\rm d}r}(r)\,\int{{\rm d}v_\theta}\,\frac{{\rm d}p}{{\rm d}v_\theta}(v_\theta|r)\,\int{{\rm d}v_r}\,\frac{{\rm d}p}{{\rm d}v_r}(v_r|r)\,\delta_{\rm D}(\tau-\tau_{\rm DF})\;.
\label{eq:merging_primordial}
\end{equation}
Since pBHs are not constantly created as stellar compact remnants, the birthrate appearing in equation (\ref{eq:merging_stellar})  is replaced here by a Dirac delta distribution $\delta_{\rm D}(\cdot)$ that select only the pBHs with dynamical friction timescale from equation (\ref{eq:tauDF}) equal to the galaxy age $\tau_{\rm DF}=\tau$.


\section{Growth of the central BH}\label{sec:seed}

The growth rate of the central BH mass due to mergers with the migrating stellar compact remnants and pBHs is given by
\begin{equation}
\dot{M}_{\bullet,\rm DF,\star/pBH}(\tau,\psi,z)=\int{{\rm d}m_\bullet}\,m_\bullet\,\frac{{\rm d}\dot{N}_{\rm DF,\star/pBH}}{{\rm d}m_\bullet}(m_\bullet,\tau|\psi,z)\; .
\end{equation}
From the growth rate, integrating over the galactic age, the contribution to the central BH mass associated to the dynamical friction process is found:
\begin{equation}
M_{\bullet,\rm DF,\star/pBH}(\tau,\psi,z)=\int_0^\tau{\rm{  d}\tau'}\,\dot{M}_{\bullet,\rm DF,\star/pBH}(\tau',\psi,z)\; .
\end{equation} 
As soon as the central BH mass has attained appreciable values,  standard gas accretion becomes increasingly efficient in growing it. For the sake of definiteness, we assume an Eddington-limited accretion with ratio $\lambda\equiv L/L_{\rm Edd}\lesssim 1$ and radiative efficiency $\eta\equiv L/\dot{M}_\bullet c^2\sim 0.1$ (see \citep{davis+11, raimundo+12, wu+13, aversa+15}), resulting in a growth rate
\begin{equation}
\dot{M}_{\bullet,\rm acc}=\frac{M_\bullet}{\tau_{\rm ef}}\; ,
\end{equation}
with $\tau_{\rm ef}=\eta/(1-\eta)\,\lambda \times (M_\bullet\,c^2/L_{\rm Edd})\simeq 4.5\times 10^7$ yr. The overall growth rate is given by the sum of the contributions from merging of stellar compact remnants and pBHs plus that from gas accretion, to yield $\dot{M}_\bullet=\dot{M}_{\bullet,\rm DF,\star}+\dot{M}_{\bullet,\rm DF,pBH}+\dot{M}_{\bullet,\rm acc}$; this can be formally integrated to find:
\begin{equation}
M_\bullet(\tau)=M_\bullet(0)e^{\tau/\tau_{\rm ef}}+\int_0^\tau\,d\tau'\,e^{(\tau-\tau')/\tau_{\rm ef}}\,(\dot{M}_{\bullet,\rm DF,\star}(\tau')+\dot{M}_{\bullet,\rm DF,pBH}(\tau'))\; .
\label{eq:central_mass}
\end{equation}

In figure \ref{fig:seed} we show the resulting growth of the central BH mass as a function of galactic age, in a galaxy at $z\sim 2$ with an average SFR $\psi\sim 300\,\rm M_\odot/yr$. The red line is the contribution from stellar compact remnants that have been funnelled toward the center via dynamical friction; it starts to become somewhat relevant at $\tau\gtrsim 10^6$ yr which is the typical lifetime of the most massive stars. Through this channel, the central BH seed can attain a mass of $\gtrsim 10^4\, M_\odot$ in a timescale of $\sim 3\times 10^7\,\rm yr$, see \citep{boco+20} for further details. 
The green patch represents the contribution to the central BH mass from migrating pBHs. The edge solid line refers to $f_{\rm pBH}\approx1$ while the dashed to $f_{\rm pBH}\approx0.01$, and the area in between to values of $f_{\rm pBH}$ within this interval. pBHs can be driven toward the nuclear region more rapidly than stellar compact remnants and hence can provide the dominant contribution to the growth of the central BH seed in its initial phases. Then the contribution from stellar compact remnants takes over
at a galactic age $\tau\gtrsim 10^6-3\times 10^7$ yr, depending on the value of $f_{\rm pBH}\approx0.01-1$. This behavior is expected since pBHs are already distributed throughout the galaxy and can immediately undergo dynamical friction, while the formation of stellar compact remnants requires some time dictated by stellar evolution processes. Nevertheless, the growth of the central BH mass by migrating pBH is rather slow, since they are distributed over all the DM halo associated to the galaxy, so that their number in the central region, where the total mass density is indeed dominated by the baryonic component, is limited. This explains why, when stellar compact remnants start to migrate toward the center, their contribution promptly overcomes that of the pBHs. These considerations are generally true but quantitatively dependent on the pBH fraction $f_{\rm pBH}$; when $f_{\rm pBH}\gtrsim 0.3$, pBHs are still able to grow the central BH seed to masses $M_\bullet\gtrsim 10^3\,M_\odot$ before other processes take over. 

The cyan patch is instead the contribution to the growth of the central BH by gas accretion, for different $f_{\rm pBH}\sim 0.01-0.1$ as above. As expected, gas accretion starts to be efficient at later times $\tau\gtrsim 3\times 10^7\,\rm yr$, when the central BH mass is $M_\bullet\gtrsim 10^4\,M_\odot$, but it rapidly grows it toward $M_\bullet\sim 10^9\,\rm M_\odot$ in a timescale of $\lesssim 300$ Myr. This is made possible thanks to the dynamical friction process, that was able to grow a heavy seed of $M_\bullet\sim 10^4-10^5\,M_\odot$ in first place, making very efficient the subsequent gas accretion onto it.
Finally, the black patch represents the sum of the three aforementioned contributions. We note that, already at intermediate times $\sim 10^7\,\rm yr$, differences in $f_{\rm pBH}$ are partly suppressed by the contribution of stellar compact remnants and at late times they are completely washed out by the last phase of nearly exponential mass growth via gas accretion. We therefore conclude that the existence of pBHs can have an important role in building up a BH seed in the very early stages. The mass of the seed so originated depends critically on the pBH mass fraction $f_{\rm pBH}$; the latter, however, cannot be probed by looking at the BH mass at late times, which is practically independent of the pBH contribution.

A final consideration regards long-living stars, i.e. stars with $m_\star<8\,\rm M_\odot$ which do not originate compact remnants, but evolve on longer lifetimes, eventually contracting to white dwarfs. In principle also these stars might be affected by dynamical friction and could sink toward the center. However, the dynamical drag acting on them is less efficient for two main reasons. First of all, the dynamical friction force scales as $F_{\rm DF}\propto m^2$ with the perturber mass, so is weaker on lighter perturbers. Second, stellar winds and energy feedback from stars not collapsed to compact objects could sweep up and and heat up the interstellar gas, making dynamical friction less effective. However, while in this treatment they are completely neglected, some of them, in fact, could reach the central regions of ETG progenitors and help in constituting a central stellar overdensity such as the nuclear star cluster. As for the central BH growth, stars in orbit around it are expected to be disrupted by the tidal effects due to the gravitational field of the BH, therefore mergers between long living stars and the central BH are not expected to happen, but they could be accreted as gas. Indeed a star with mass $m_\star$ and radius $r_\star$ is tidally disrupted by a BH of mass $M_\bullet$ when it reaches the tidal radius $r_T=r_\star\,(M_\bullet/m_\star)^{1/3}$. In order to not be disrupted and contribute to the merging rates with the central BH the tidal radius should be larger than the Schwarzschild radius $r_T>2\,G\,M_\bullet/c^2$, implying, for a star as the sun, $M_\bullet\gtrsim 10^8\,\rm M_\odot$, which is not the case for the initial stages when the proposed mechanism is relevant.

\section{GW emission and detection}\label{sec:gw}

The process of migration via dynamical friction and merging with the central BH mass can be probed through GW observations (see also \citep{boco+20}). Equations (\ref{eq:merging_stellar}) and (\ref{eq:merging_primordial}) represents, respectively, the merging rates per unit mass $m_\bullet$ of migrating stellar compact remnants and pBHs  with the central mass $M_\bullet$. Since the strength of the GW signal is proportional to the chirp mass $\mathcal{M}_{\bullet\bullet}\equiv(M_\bullet\,m_\bullet)^{3/5}/(M_\bullet+m_\bullet)^{1/5}$, it is convenient to change variable in equations (\ref{eq:merging_stellar}) and (\ref{eq:merging_primordial}) using the relation between galactic age $\tau$ and the central BH mass $M_\bullet(\tau|z,\psi)$, 
set by equation (\ref{eq:central_mass}). Thus we obtain the merging rate at time $\tau$ per unit chirp mass $\mathcal{M}_{\bullet\bullet}$ in a galaxy with average SFR $\psi$ at redshift $z$:  
\begin{equation}
\frac{{\rm d}\dot{N}_{\rm DF, \star/pBH}}{{\rm d}\mathcal{M}_{\bullet\bullet}}(\mathcal{M}_{\bullet\bullet},\tau|\psi,z)=\frac{{\rm d}\dot{N}_{\rm DF, \star/pBH}}{{\rm d}m_\bullet}(m_\bullet(\mathcal{M}_{\bullet\bullet},M_\bullet(\tau)),\tau|\psi,z)\,\frac{{\rm d }m_\bullet}{{\rm d}\mathcal{M}_{\bullet\bullet}}(\mathcal{M}_{\bullet\bullet},\tau,\psi,z)\; .
\label{eq:dotNdmchirp}
\end{equation}

The overall cosmological merging rates at redshift $z$ (or cosmic time $t_z$) per unit chirp mass can be derived from equation (\ref{eq:dotNdmchirp}) as
\begin{equation}
\begin{split}
\frac{{\rm d}\dot{N}_{\rm DF, \star/pBH}}{{\rm d}V\,{\rm d}\mathcal{M}_{\bullet\bullet}}(\mathcal{M}_{\bullet\bullet},z) &=\int{{\rm d}\psi}\,\frac{{\rm d}N}{{\rm d}V\,{\rm d}\psi}(\psi,z)\, \int_{t_z-\tau_\psi}^{t_z}{{\rm d}t_{\rm form}}\, \frac{{\rm d}p}{{\rm d}t_{\rm form}}(t_{\rm form}|\psi)\times\\
\\
&\times\frac{{\rm d}\dot{N}_{\rm DF, \star/pBH}}{{\rm d}\mathcal{M}_{\bullet\bullet}}(\mathcal{M}_{\bullet\bullet},t_z-t_{\rm form}|\psi,z)\,\Theta(t_z-t_{\rm form}\lesssim t_{\rm max})\; .
\end{split}
\label{eq:dotNdVdmchirp}
\end{equation}
Here the contribution of different galaxies is weighted by the SFR function ${\rm d}N/{\rm d}V/{\rm d}\psi$, expressing the number density of galaxies per logarithmic bin of SFR and comoving volume at different redshifts. This statistics has been derived from a combination of the observed UV/IR/(subm-)mm/radio luminosity functions by \citep{mancuso+16a}, has been validated against a wealth of independent datasets (see \citep{mancuso+16a, mancuso+16b, mancuso+17, lapi+17a, lapi+17b}) and has been extensively used in numerous previous works (see e.g. \citep{boco+19, boco+20, boco+21, scelfo+20, capurri+21}). Moreover, $t_{\rm form}$ is the formation time of the galaxy and ${\rm d}p/{\rm d}t_{\rm form}$ is the related probability distribution, that we take flat for simplicity. Other relevant quantities are the star formation timescale $\tau_\psi$, that we take following the lines of \citep{boco+19} and is of the order of $\tau_\psi\simeq$ some $10^8\,\rm yr$, and the maximum time $t_{\rm max}$ over which the dynamical friction process is active, as specified in terms of the Heaviside step function $\Theta(\cdot)$. 

The above equation can be understood along the following lines. The merging rates in a galaxy with given SFR $\psi$ at redshift $z$ are first computed at the galactic age $\tau=t_z-t_{\rm form}$, averaging over all the possible formation times. Then the result is summed over all the galaxies with different SFR, weighted by their statistics. 
The meaning of the star formation timescales $\tau_\psi$ and of the activity timescale $t_{\rm max}$ for dynamical friction is more subtle. It is well established that, when the central BH mass has grown to substantial values $M_\bullet\gtrsim$ some $10^8\, M_\odot$, feedback in the form of energy and momentum from the active nucleus will affect the host galaxy, removing gas and quenching the star formation; this typically occurs on a timescale $\tau_\psi$ which depends on the SFR and redshift, but typically amounts to several $10^8$ yr. At this point, the  gas reservoir is substantially reduced even on large scales and the dynamical friction process is also stopped, to imply $t_{\rm max}\sim \tau_\psi$. This will be our fiducial choice here, but we caveat that the BH feedback can be effective in depleting the gas reservoir from the nuclear regions, even before the galaxy-wide SFR is quenched, to imply $t_{\rm max}\lesssim \tau_\psi$; a strict lower limit to $t_{\rm max}$ could be some $10^7$ yr, which is the timescale when the dynamical friction contribution to the growth of the central seed mass becomes subdominant with respect to gas accretion (see Appendix).  

Integrating equation (\ref{eq:dotNdVdmchirp}) over the chirp mass we compute the cosmic merging rate density due to the dynamical friction process both for stellar compact remnants and for pBHs. We show the outcome as a function of redshift in figure \ref{fig:mergingrate}. The contribution of stellar compact remnants, which does not depend on $f_{\rm pBH}$, is shown as a solid red line. Their number density is comparable to that for the merging of isolated BH binaries, at least according to some literature estimates. The pBH contribution is instead represented by the green patch for $f_{\rm pBH}$ ranging between $0.01$ (dashed green edge line) and $1$ (solid green edge line). Plainly, the number density of pBHs migrating toward the center grows with $f_{\rm pBH}$ but is always substantially lower with respect to that of stellar compact remnants.

The chirp mass distribution ${\rm d}\dot{N}_{\rm DF, \star/pBH}/{\rm d}V/d\mathcal{M}_{\bullet\bullet}$ of equation (\ref{eq:dotNdmchirp}) at a reference redshift $z\sim 2$ is shown in figure \ref{fig:mergingrate_chirp}. The red patch refers to stellar compact remnants, while the green patch to pBHs; solid lines are for $f_{\rm pBH}\approx 1$ and dashed lines for $f_{\rm pBH}\approx 0.01$. We notice that while the overall number density of stellar compact remnants mergers is independent of $f_{\rm pBH}$ (see figure \ref{fig:mergingrate}), their chirp mass distribution instead does depend on it. In fact, as seen in section \ref{sec:seed}, pBHs migrate toward the galactic center earlier than stellar compact remnants and contribute to the initial growth of the central BH seed. A larger number of pBHs implies a faster growth of the central BH in the initial stages, so increasing the chirp mass of the subsequent merging events with the stellar compact remnants. This explains why the chirp mass distribution for stellar compact remnants shifts towards higher chirp masses for larger values of $f_{\rm pBH}$. Notice that the in the low chirp mass range $\mathcal{M}_{\bullet\bullet}\lesssim 500\,\rm M_\odot$ the distribution presented is scarcely affected by the value of the Eddington ratio $\lambda$ since gas accretion starts to become important at later stages, when the central BH already attained a mass $M_\bullet\gtrsim 10^4\,\rm M_\odot$ and the typical chirp masses for mergers are higher.

We now compute the detected GW event rate per redshift and chirp mass bin, following the procedure outlined in \citep{boco+20}, to obtain 
\begin{equation}
\begin{split}
\frac{{\rm d}\dot{N}_{\rm GW}}{{\rm d}z\,{\rm d}\mathcal{M}_{\bullet\bullet}}(z,>\rho_0)=&\frac{1}{1+z}\,\frac{{\rm d}V}{{\rm d}z}\,\frac{{\rm d}\dot{N}_{\rm DF}}{{\rm d}V{\rm d}\mathcal{M}_{\bullet\bullet}}(\mathcal{M}_{\bullet\bullet}, z)\,\int{{\rm d}q}\,\frac{{\rm d}p}{{\rm d}q}(q|\mathcal{M}_{\bullet\bullet},z)\times\\
\\
&\times\int d\Delta t_{\rm obs}\frac{dp}{d\Delta t_{\rm obs}}\Theta[\bar\rho(\mathcal{M}_{\bullet\bullet}, q, \Delta t_{\rm obs}, z)\gtrsim \rho_0]\; ,
\end{split}
\label{eq:detection}
\end{equation}
where ${\rm d}V/{\rm d}z$ is the differential comoving volume, $q\equiv m_\bullet/M_\bullet$ is the mass ratio between the migrating remnant and the central BH, ${\rm d}p/{\rm d}q$ is the related probability distribution, $\Delta t_{\rm obs}$ is the overall observation time, $\bar\rho$ is the sky-averaged SNR of the event dependent on redshift, chirp mass, mass ratio and observational time and $\rho_0$ is the SNR threshold selected for the detection; we use $\rho_0=8$ for ET and DECIGO and $\rho_0=30$ for LISA. The probability distribution of the mass ratio $q$ can be computed as:
\begin{equation}
\frac{{\rm d}p}{{\rm d}q}(q|\mathcal{M}_{\bullet\bullet},z)\propto\int{{\rm d}\psi}\, \frac{{\rm d}N}{{\rm d}\psi\,{\rm d}V}(\psi,z)\, \frac{{\rm d}N_{\rm DF}}{{\rm d}m_\bullet}(m_\bullet(\mathcal{M}_{\bullet\bullet}, q),\tau(\mathcal{M}_{\bullet\bullet}, q)|\psi,z)\, \frac{{\rm d}m_{\bullet}}{{\rm d}q}
\end{equation}
normalized such that $\int{{\rm d}q}\,{\rm d}p/{\rm d}q=1$. The sky-averaged SNR of the GW event is
\begin{equation}
\bar\rho=\frac{G^{5/6}}{\sqrt{3}\pi^{2/3}c^{3/2}}\,\frac{[(1+z)\mathcal{M}_{\bullet\bullet}]^{5/6}}{D_L(z)}\,\sqrt{\int_{f_{\rm in}}^{f_{\rm isco}}\frac{{\rm d}f}{f^{7/3}\,S(f)}}\; ,
\end{equation}
where $D_L$ is the cosmological luminosity distance, $S(f)$ is the total sensitivity curve of the detector, $f_{\rm isco}$ is the GW redshifted frequency of the innermost stable circular orbit and $f_{\rm in}$ is a lower limit that takes into account the frequency evolution over the observational time $\Delta t_{\rm obs}$. The quantity $f_{\rm isco}$\footnote{Actually for the computation of the detection rates for ET and DECIGO we also include the merger and ringdown phases, treated as in \citep{boco+19}} can be computed as
\begin{equation}
f_{\rm isco}\simeq\frac{4400}{1+z}\left(\frac{M_{\rm tot}}{M_\odot}\right)^{-1}\,\rm Hz\; ,
\end{equation}
where $M_{\rm tot}\equiv m_\bullet+M_\bullet=\mathcal{M}_{\bullet\bullet}(1+q)^{6/5}q^{-3/5}$ is the total mass of the merging objects. For high-frequency interferometers like ET and DECIGO the frequency shift is very rapid, so that $f_{\rm in}$ is well outside the observational window of the instrument and the approximation $f_{\rm in}\simeq 0$ applies. Therefore the SNR $\bar\rho$ in equation (\ref{eq:detection}) is independent on $\Delta t_{\rm obs}$ and the related integration does not matter any longer. Contrariwise, for LISA the frequency shift is rather slow and $f_{\rm in}$ can be determined by integrating the orbital-averaged equations (see \citep{peters+64}), to get 
\begin{equation}
f_{\rm in}\simeq f_{\rm isco}\left[1+\frac{1}{5}\left(\frac{2}{3}\right)^4\frac{q^{8/5}}{(1+q)^{16/5}}\frac{c^3\Delta t_{\rm obs}}{G\mathcal{M}_{\bullet\bullet}(1+z)}\right]^{-3/8}\; .
\end{equation}

Integrating equation (\ref{eq:detection}) over the chirp mass yields the detected GW rates as a function of redshift for stellar compact remnants and pBHs. The results for ET, DECIGO and LISA are shown in figure \ref{fig:detected_all}. All in all, the outcomes mirror the intrinsic merging rates (see figure \ref{fig:mergingrate}). In the ET case (top left panel) we stress that a larger value of $f_{\rm pBH}$ increases the detected GWs from pBHs and correspondingly decrease those from stellar compact remnants. 
This is again due to the fact that pBHs contribute mainly to the initial growth of the central BH mass. If the number of pBHs increase, the central mass grows faster, and the migrating stellar compact remnants will tend to merge with an already massive central BH seed; this will in turn make the GW signal to exit the ET observational window. This can also be seen from figure \ref{fig:mergingrate_chirp}, in that for larger $f_{\rm pBH}$ the number of merging stellar compact remnants is the same, but their chirp mass distribution is shifted towards higher masses, so reducing the number of events detectable by ET. The ET detection rates of stellar compact remnants and pBHs are of the same order for $f_{\rm pBH}\sim 0.3$. We stress that, for any value of $f_{\rm pBH}$, the number of detected events from the dynamical friction process are lower than those associated to isolated compact binary mergers, as can be seen from the dotted black line in the top panel of figure \ref{fig:detected_all}. 

To disentangle the events related to dynamical friction, one possibility is to rely on those with chirp masses much larger than expected from isolated binary mergers.
To this purpose, in figure \ref{fig:detected_all} (top right panel) we also plot the rates of GW signals associated to the dynamical friction process with chirp mass $\mathcal{M}_{\bullet\bullet}\gtrsim 200\, M_\odot$. Plainly the detected rates are reduced somewhat, especially the ones for stellar compact remnants with $f_{\rm pBH}\sim 0.01$, but
their number is still sizeable; a detection of these high-chirp mass event could probe the actual occurrence of the dynamical friction process invoked in this work.
For DECIGO (bottom left panel) and LISA (bottom right panel), although the detected GW rates from pBH mergers are still strongly dependent on $f_{\rm pBH}$, those from stellar compact remnants are not. This is because the GW events entering in the DECIGO and LISA observational band are extreme mass ratio inspirals between a migrating compact remnant and an already large central BH mass $10^5-10^8\,M_\odot$. This occurs at galactic ages when the growth of the central BH is mainly dominated by migrating stellar compact remnants or gas accretion; thus the possible effects of pBHs, and the related dependence on $f_{\rm pBH}$, is almost completely washed out (see also figure \ref{fig:seed}).

Figure \ref{fig:detected_chirp} shows the contribution to the detected rates at $z\sim 2$ from different chirp masses, associated to stellar compact remnants (top panel) and pBHs (bottom panel). The black patch represents the intrinsic chirp mass distribution, while the blue, green and orange patches refer to the chirp mass distribution detected by ET, DECIGO and LISA, respectively; the edge lines refer to $f_{\rm pBH}\approx1$ (solid) and to $f_{\rm pBH}\approx0.01$ (dashed). ET and LISA are almost complementary: ET will detect events with chirp mass $\mathcal{M}_{\bullet\bullet}\sim 10-500\,M_\odot$ corresponding to central BH masses up to $M_\bullet\sim 10^5\,\rm M_\bullet$ and occurring in the initial stages of the seed growth; LISA will detect events with chirp mass $\mathcal{M}_{\bullet\bullet}\sim 1000-5000\,M_\odot$ corresponding to central BH masses $M_\bullet\sim 10^5-10^8\,M_\odot$ and occurring in the late stages of central BH growth.
On the other hand, the large frequency band and exquisite sensitivity of DECIGO will allow to probe the intrinsic chirp mass distribution up to a chirp mass $\mathcal{M}_{\bullet\bullet}\lesssim 5000\, M_\odot$, with incredibly high detection rates. 

\section{Stochastic GW background}\label{sec:sgwb}

The stochastic gravitatitonal wave background is originated by the incoherent superposition of undetected GW signals, and can be computed as
\begin{equation}
\begin{split}
\Omega_{\rm GW}(f_{\rm obs}) & =\frac{8\pi G\,f_{\rm obs}}{3\,H_0^3\,c^2}\int\frac{{\rm d}z}{(1+z)\,E(z)}\int {\rm d}\mathcal{M}_{\bullet\bullet}\, \frac{{\rm d}\dot{N}}{{\rm d}V{\rm d}\mathcal{M}_{\bullet\bullet}}\,\int {\rm d}q\frac{{\rm d}p}{{\rm d}q}(q|\mathcal{M}_{\bullet\bullet},z)\times\\
\\
&\times\frac{{\rm d}E}{{\rm d}f}(f(f_{\rm obs},z)|\mathcal{M}_{\bullet\bullet}, q)\int {\rm d}\Delta t_{\rm obs}\, \frac{{\rm d}p}{{\rm d}\Delta t_{\rm obs}}\, \Theta[\bar\rho(\mathcal{M}_{\bullet\bullet}, q, \Delta t_{\rm obs}, z)\lesssim \rho_0]\; ,
\end{split}
\end{equation}
where ${\rm d}E/{\rm d}f$ is the emitted GW energy spectrum (see e.g. \citep{zhu+11, boco+19}), depending on the chirp mass $\mathcal{M}_{\bullet\bullet}$ and on the mass ratio $q$ of the merging objects; the Heaviside function in the innermost integral ensures the summation over the unresolved signals with SNR  $\bar\rho<\rho_0$. The SGWB generated by all events, resolved and unresolved, is obtained by setting the detection threshold $\rho_0\rightarrow\infty$, so that the innermost integral goes to $1$.

In figure \ref{fig:SGWB_ET} we show the total SGWB originated from both resolved and unresolved events; the orange patch is for stellar compact remnants and the blue patch for pBHs. Solid lines are for $f_{\rm pBH}\approx1$ while dashed ones for $f_{\rm pBH}\approx0.01$. Plainly, the contribution to the SGWB from pBHs increases at higher $f_{\rm pBH}$ as the number of merging events is larger. On the other hand, the contribution from stellar compact remnants shifts towards lower frequencies as $f_{\rm pBH}$ increases. This reflects the shift at higher chirp masses of the stellar compact object merger rates, see discussion in  section \ref{sec:gw}. The red and blue patches, instead, represent the residual SGWB from stellar compact remnants and pBHs when the resolved events by ET are subtracted. This originates a sharp drop at $f\gtrsim 1$ Hz, where the ET starts detecting almost all the events, subtracting them from the unresolved background. For comparison, the SGWB originated by the merging of isolated BH-BH binaries is also plotted as a dotted grey line (e.g., \citep{boco+19}). We note that the range of frequencies involved for the two processes is rather different, with $10^{-6}\lesssim f [{\rm Hz}]\lesssim 10$ for the dynamical friction induced mergers described in this paper and $10^{-2}\lesssim f [{\rm Hz}]\lesssim 10^4$ for the merging of isolated BH-BH binaries. As a matter of fact, since ET would be able to resolve almost all the events falling in its frequency sensitivity window, it would be very unlikely for it to detect the SGWB coming from the dynamical friction process.

In figure \ref{fig:SGWB_DECIGO}, we show the corresponding results on the SGWB for DECIGO.
The residual SGWB background of unresolved events is sharply truncated with respect to the total  for frequencies $f [{\rm Hz}]\gtrsim 10^{-2}$, corresponding to the DECIGO sensitivity band. We also plot as black solid line the DECIGO sensitivity curve to the background (see \citep{moore+15}). The exquisite sensitivity to the background for DECIGO will allow to characterize the SGWB at frequencies  $f [{\rm Hz}]\lesssim 10^{-2}$; this will be more easily achieved for the background originated by migrating stellar BHs, though also that from migrating pBHs can be detected, especially if $f_{\rm pBH}\lesssim 1$.

In figure \ref{fig:SGWB_LISA} we show the corresponding results on the SGWB for LISA. The residual SGWB background of unresolved events is reduced with respect to the total 
for frequencies $10^{-3}\lesssim f [{\rm Hz}]\lesssim 1$, corresponding to the LISA sensitivity band. However, the reduction is not as sharp as for ET since, as shown also in figure \ref{fig:detected_chirp}, LISA will not detect all the events occurring in its sensitivity band. We also plot as black solid line the LISA sensitivity curve to the background (see \citep{robson+19}). The SGWB from the dynamical friction process is fully detectable with LISA (both for migrating stellar and pBHs).

We caveat that the above predictions concerning the SGWB might be affected by many uncertainties. For example, as already discussed in section \ref{sec:gw}, the duration of the dynamical friction process could be shorter, lowering the number of mergers and the amplitude of the SGWB. On the other hand, the presence of compact objects binaries from stellar or primordial origin, could somewhat lower the dynamical friction timescales, increasing the number of compact objects merging with the central BH and making the SGWB stronger.  Given that, our prediction is remarkable, because no other astrophysical mechanisms can originate such a strong background in this frequency range; a future detection of this could represent a smoking gun to test the occurrence of the dynamical friction and its role in the BH seed growth.  

\section{Conclusions}\label{sec:conclusion}

We have investigated the issue of seed formation and growth in dusty star forming galaxies, relying on and extending the framework proposed by \citep{boco+20}.
The latter envisages the migration of stellar compact remnants (neutron stars and stellar-mass black holes) via gaseous dynamical friction towards the nuclear galaxy region and their subsequent merging to grow a massive central black hole (BH) seed. Specifically, in the present paper we have included primordial BHs (pBHs) as an additional component participating in the seed growth process. Moreover, we have predicted the stochastic gravitational wave background originated during the seed growth, both from stellar compact remnant and from pBH mergers.

After a brief recap of the gaseous dynamical friction process and a short description of the modeling setup (section \ref{sec:DF}), we have analyzed (section \ref{sec:seed}) the growth of the central BH mass as a function of the galactic age for different pBH-to-DM fraction $f_{\rm pBH}$. We have found that the contribution of pBHs for $f_{\rm pBH}\approx0.01-1$ can be effective in growing a central seed of mass $\sim 10^3-10^4\,M_\odot$ at early galactic ages $\tau\lesssim 10^6-3\times 10^7\,\rm yr$. 
The pBH contribution is then overwhelmed by that from migrating stellar compact remnants, which reach later the nuclear regions but are far more numerous. 
We have pointed out that differences in the central BH mass due to the value of $f_{\rm pBH}$ are first smoothed out by the contribution of migrating stellar compact remnants and then washed out by the subsequent phase of gas accretion onto the formed central massive BH. Therefore, the value of $f_{\rm pBH}$ cannot be probed by looking at the relic value of the central (super)massive BH mass. 

We emphasize that the above results are based on an analytical treatment of a rather complex process that cannot keep into account all the detailed concomitant and co-spatial physical mechanisms occurring in the central regions of dusty star-forming galaxies such as gas clumpiness, supernova feedback, natal kicks, three or many body encounters and exact compact objects velocity structure. Nevertheless, a back of the envelope estimation of these effects has been provided throughout the work (see also section 5 of \citep{boco+20}) as a preliminary demonstration of the robustness of the investigated process. Further investigation of the details of the dynamical friction process can only be achieved via a full hydrodynamical simulation at high spatial resolution which is beyond the scope of the present work.

We have then computed the rate of GW emission during the seed growth process, and the related detection rate with future interferometers like ET, DECIGO and LISA. We have shown that, while the intrinsic rate of emitted GWs is always smaller for pBHs with respect to stellar compact remnants, the detected rate could strongly depend on $f_{\rm pBH}$. In the ET case, for small values of $f_{\rm pBH}\lesssim 0.3$ the detected rates from pBHs are much less than stellar compact remnant ones. However, increasing $f_{\rm pBH}$ originates a shift in the chirp masses of stellar compact remnant events, decreasing their detectable fraction. As a consequence, for $f_{\rm pBH}\gtrsim 0.3$ GW detected events from pBHs outnumber the ones associated with stellar compact remnants. In the case of DECIGO and LISA, instead, this effect is much weaker and the detection rates are comparable only for an extreme value $f_{\rm pBH}\sim 1$. We stress that the optimal frequency band and exquisite sensitivity of DECIGO will allow to probe the intrinsic chirp mass distribution of the merger events induced by dynamical friction migration (both for stellar remnants and pBHs) over an extended chirp mass range.

Finally, we have computed the stochastic GW background originated during the growth of the central BH by the merger events of migrating stellar compact remnants and pBHs (see section \ref{sec:sgwb}). We have highlighted that the background extends over a wide range of frequencies $10^{-6}\lesssim f [{\rm Hz}]\lesssim 10$, which is very different from that associated to isolated merging compact binaries. The detection of such a background could be challenging for ET but within the reach of DECIGO and LISA. All in all, it would constitute a smoking gun to confirm the proposed process of BH seed growth. We further stress that the characterization of possible astrophysical sources contributing to the background will be fundamental toward the challenging search for a primordial contribution of cosmological origin. 

\acknowledgments
This work is partially supported by the PRIN MIUR 2017 prot. 20173ML3WW 002, `Opening the ALMA window on the cosmic evolution of gas, stars and supermassive black holes' and by the EU H2020-MSCA-ITN-2019 Project 860744 `BiD4BEST: Big Data applications for Black hole Evolution STudies'. We thank the anonymous referee for valuable comments, which led to a deeper analysis and to a substantial improvement of the manuscript. L.B. thanks Samuele Campitiello for helpful discussions.

\begin{appendix}

\section{Impact of the dynamical friction duration}

In section \ref{sec:gw} we have discussed the role of the duration $t_{\rm max}$ over which the dynamical friction process is active. Throughout the paper we have shown results for our fiducial assumption $t_{\rm max}\sim\tau_\psi$, i.e., a dynamical friction process extending over the star formation duration in the host. However, energy/momentum feedback from the central BH could be effective in depleting the gas reservoir from the nuclear regions even before the galaxy-wide SFR is quenched, implying $t_{\rm max}\lesssim \tau_\psi$. As discussed in the text, a reasonable lower limit $t_{\rm max}\approx 5\times 10^7\,\rm yr$ applies; in this appendix we discuss how the main results of the present paper are affected. 

\begin{figure*}
    \centering
    \includegraphics[width=1.\textwidth]{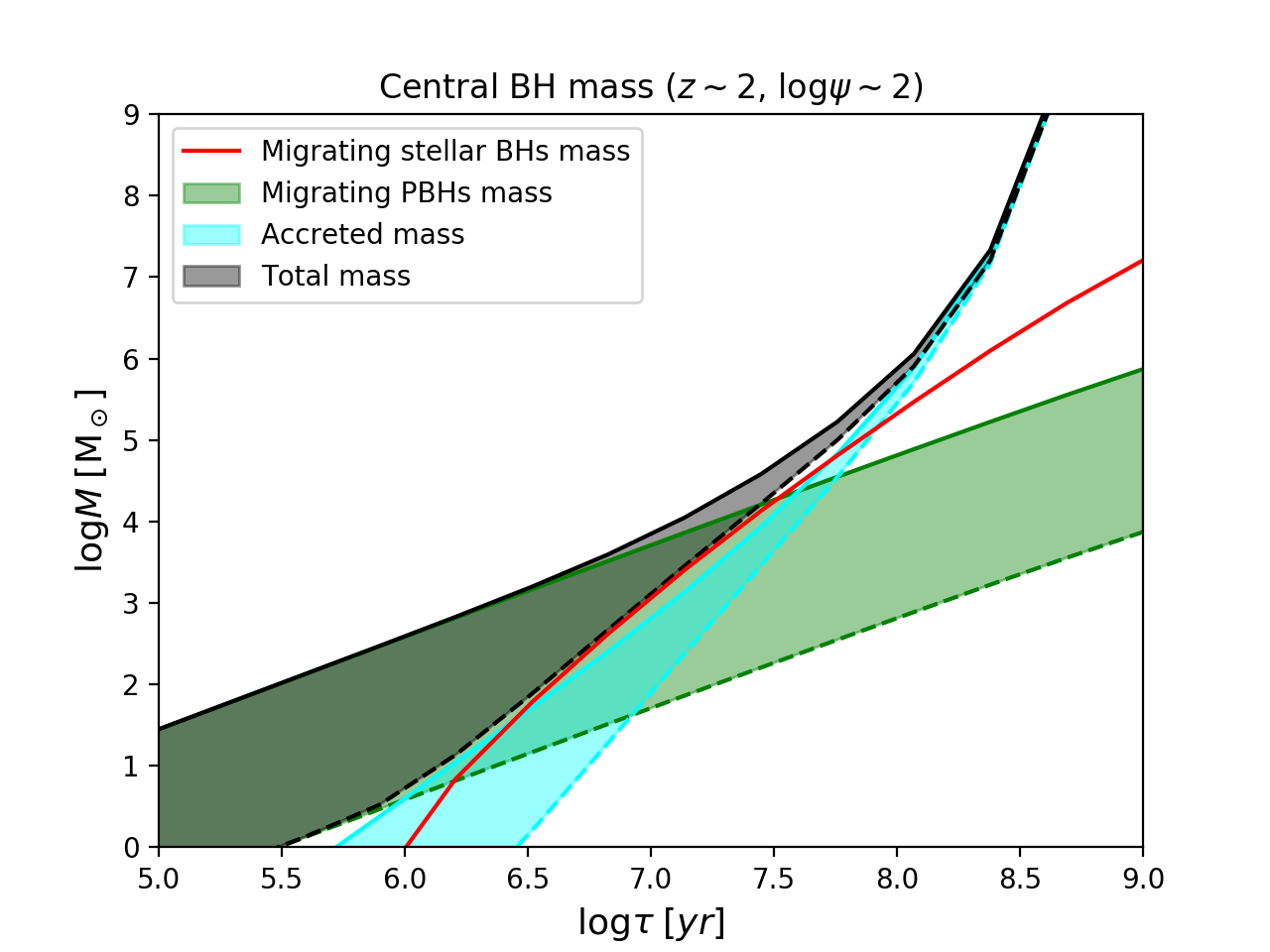}
    \caption{Growth of the central BH mass $M_\bullet$ as a function of the galactic age $\tau$, due to the gaseous dynamical friction process discussed in this paper. For the sake of definiteness a galaxy with average SFR $\psi\sim 300\, M_\odot$ yr$^{-1}$ at redshift $z\approx 2$ has been considered. Red line illustrates the contribution to the growth from migrating stellar compact remnants, green shaded area from migrating pBHs, cyan shaded area from gas accretion onto the central BH, and black shaded area refers to the the total. Both shaded areas show the effect of varying the pBH-to-DM fraction $f_{\rm pBH}$ from $0.01$ (dashed edge lines) to $1$ (solid edge lines).}
    \label{fig:seed}
\end{figure*}

\begin{figure*}
    \centering
    \includegraphics[width=1.\textwidth]{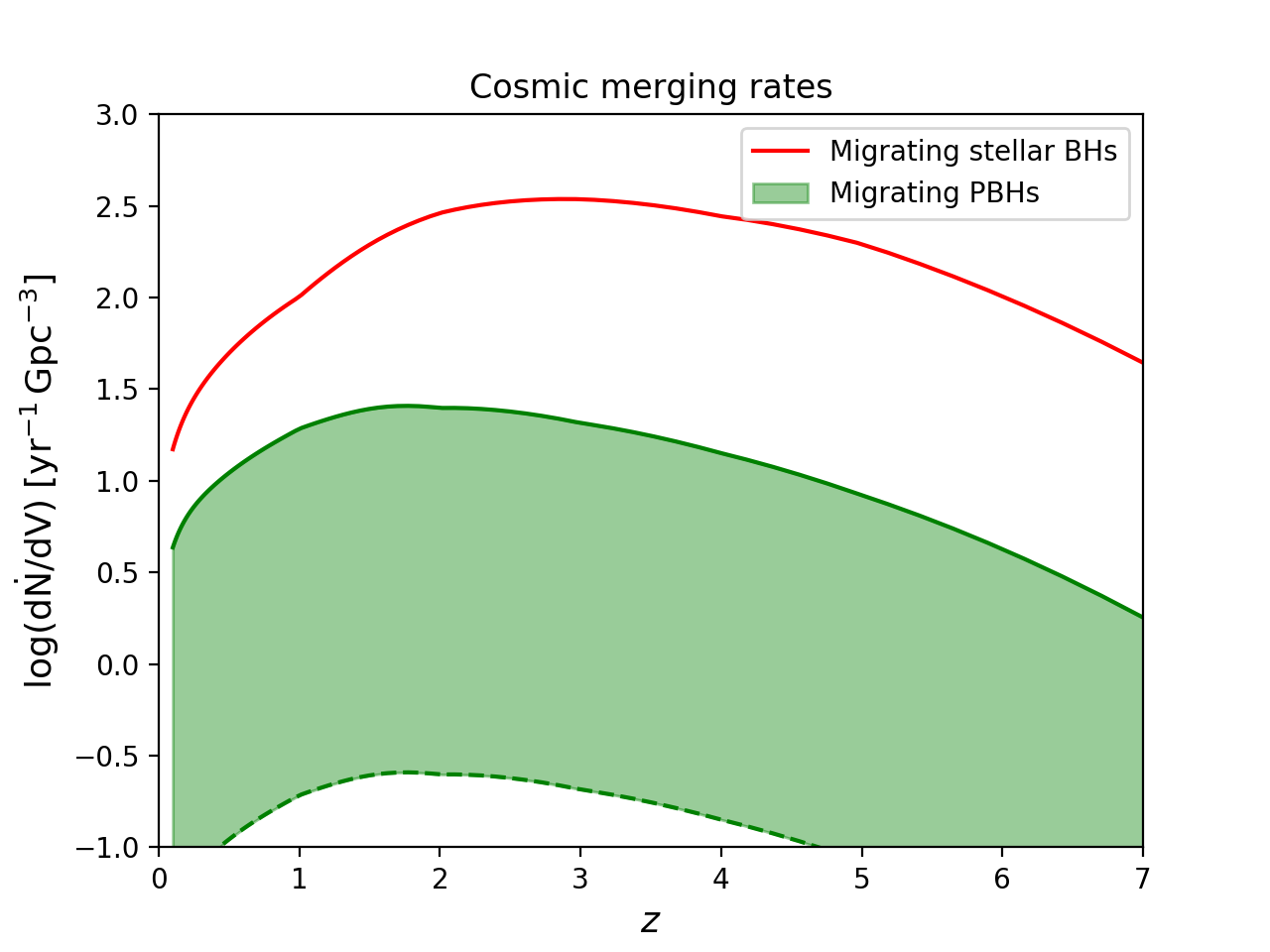}
    \caption{Cosmic merging rate density as a function of redshift, due to the gaseous dynamical friction process discussed in this paper. Red solid line refers to migrating stellar compact remnants, green shaded area to migrating pBHs, for different pBH-to-DM fraction $f_{\rm pBH}$ ranging from $0.01$ (dashed edge line) to $1$ (solid edge line).}
    \label{fig:mergingrate}
\end{figure*}

\begin{figure*}
    \centering
    \includegraphics[width=1.\textwidth]{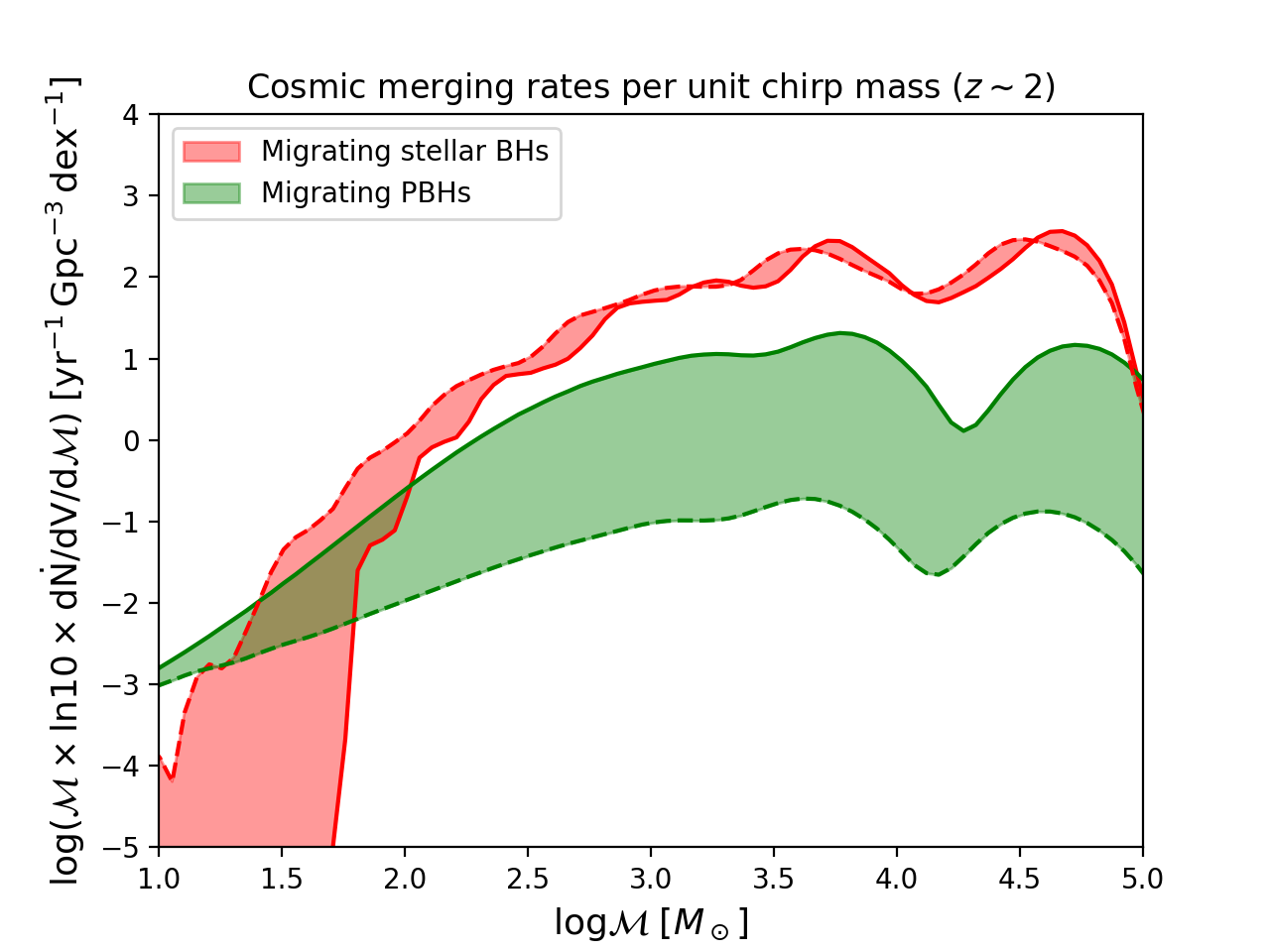}
    \caption{Cosmic chirp mass distribution at a reference redshift $z\sim 2$. Red shaded area refers to migrating stellar compact remnants, green shaded area to migrating pBHs, for different pBH-to-DM fraction $f_{\rm pBH}$ ranging from $0.01$ (dashed edge lines) to $1$ (solid edge lines). }
    \label{fig:mergingrate_chirp}
\end{figure*}

\begin{figure*}
    \centering
    \includegraphics[width=1.\textwidth]{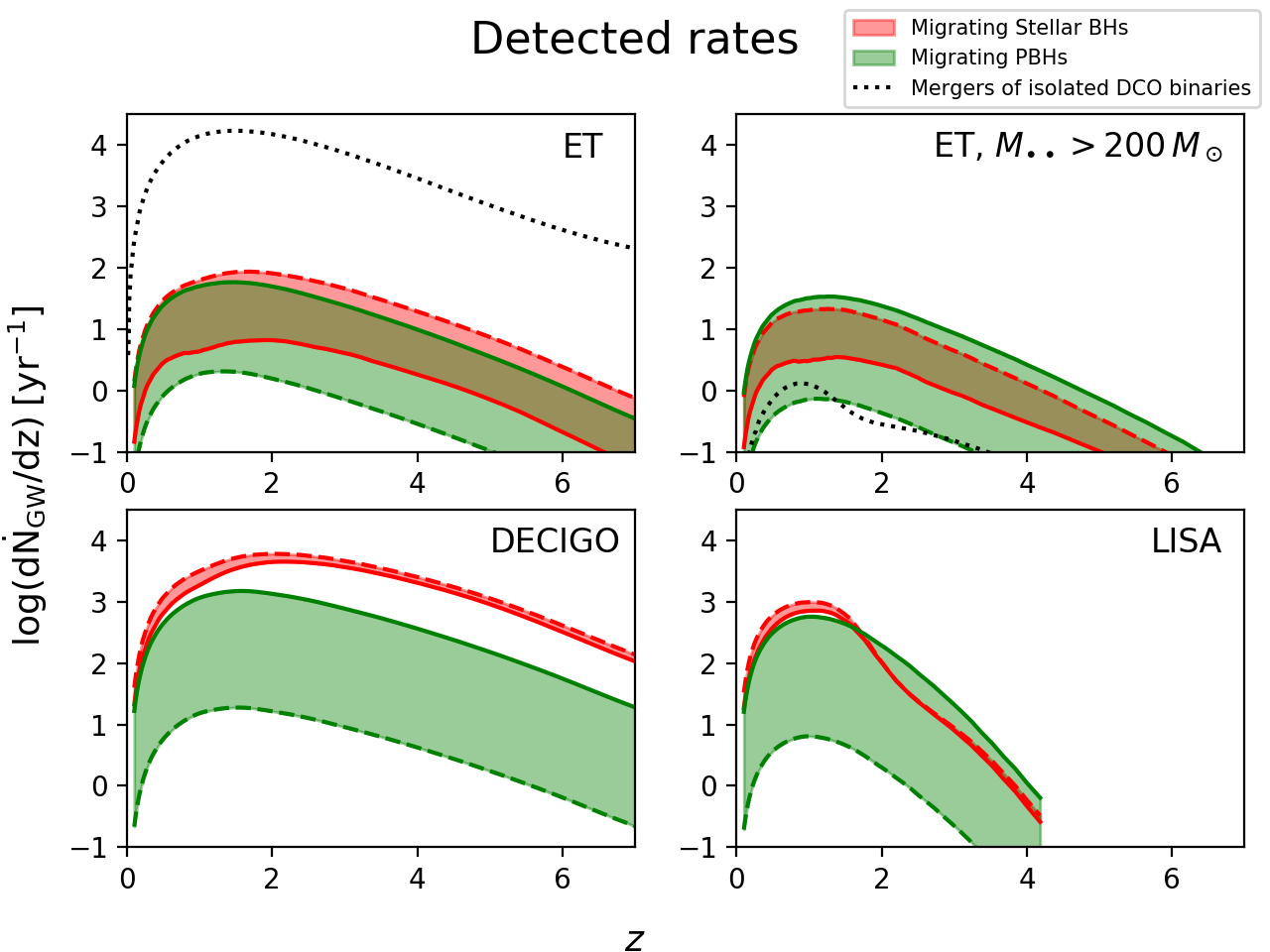}
    \caption{GW detected event rates for ET (top left panel), for ET and chirp masses $\mathcal{M}_{\bullet\bullet}>200\,\rm M_\odot$ (top right panel), DECIGO (bottom left panel) and LISA (bottom right panel). 
    A signal-to-noise ratio $\rho>8$ is adopted for ET and DECIGO, while $\rho>30$ is adopted for LISA. Red shaded areas refer to migrating stellar compact remnants, green shaded areas to migrating pBHs, for different pBH-to-DM fraction $f_{\rm pBH}$ ranging from $0.01$ (dashed edge lines) to $1$ (solid edge lines). In the top panels, the dotted line illustrates the detection rates for ET associated to mergers of isolated double compact objects binaries.}
    \label{fig:detected_all}
\end{figure*}

\begin{figure*}
    \centering
    \includegraphics[width=0.7\textwidth]{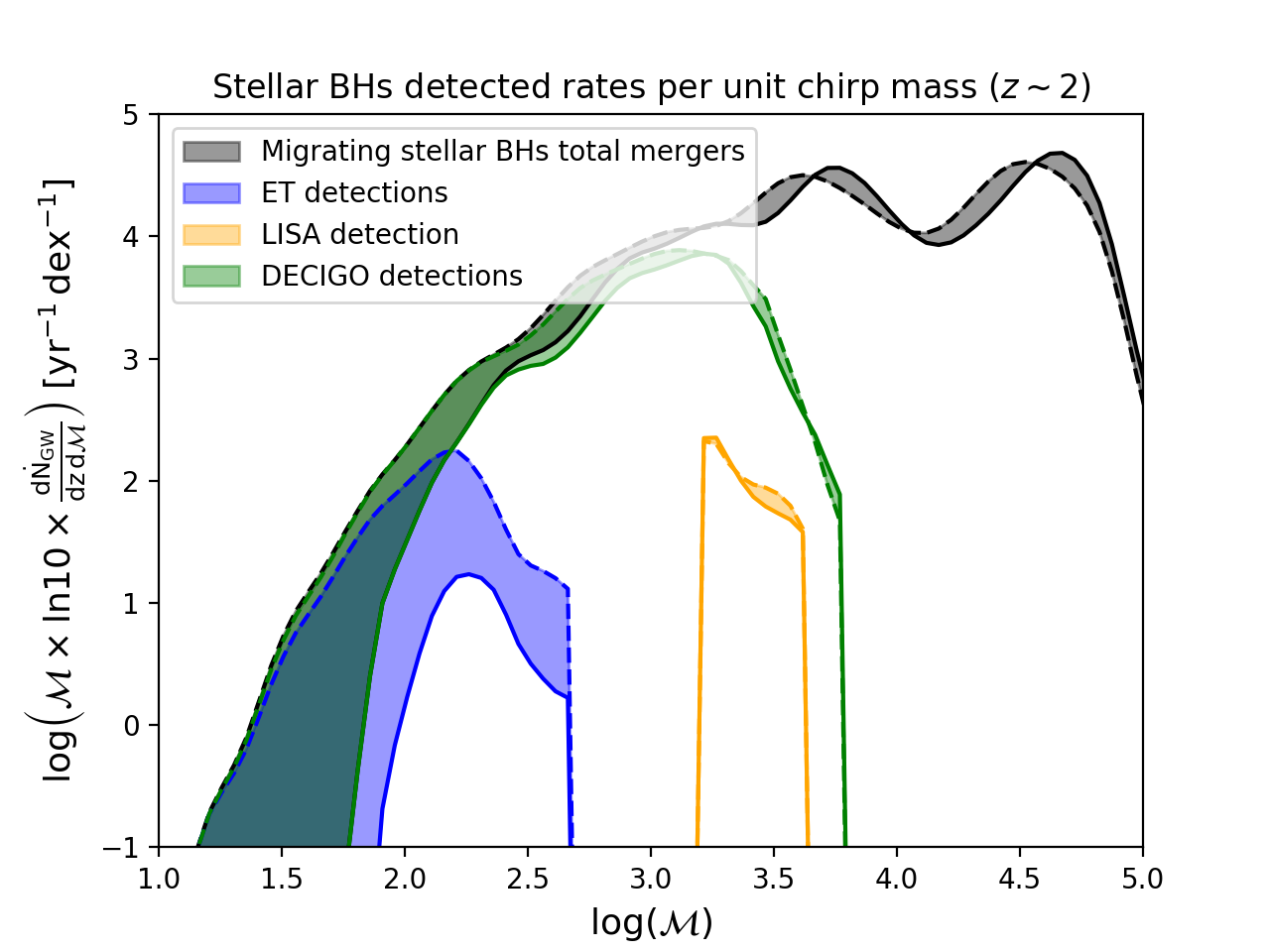}
    \includegraphics[width=0.7\textwidth]{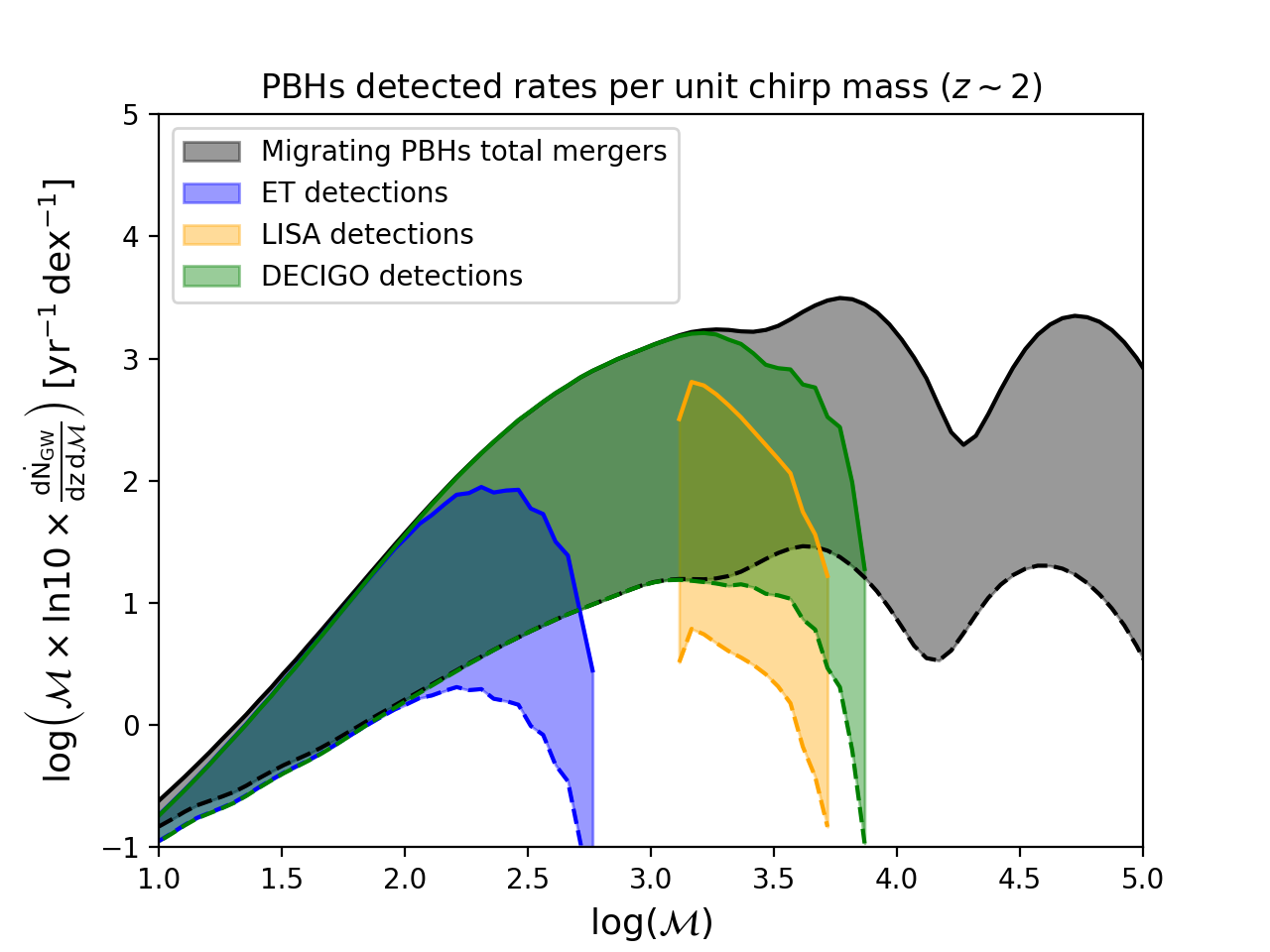}
    \caption{Chirp mass distribution of detected events at $z\sim 2$ for migrating stellar compact remnants (top panel) and for migrating pBHs (bottom panel). Grey shaded areas show the intrinsic distribution of all events independent of their detection, blue shaded areas refers to ET detections, yellow shaded areas to LISA detections and green shaded areas to DECIGO detections. The shaded areas show the effect of varying the pBH-to-DM fraction $f_{\rm pBH}$ from $0.01$ (dashed edge lines) to $1$ (solid edge lines). }
    \label{fig:detected_chirp}
\end{figure*}

\begin{figure*}
    \centering
    \includegraphics[width=1.\textwidth]{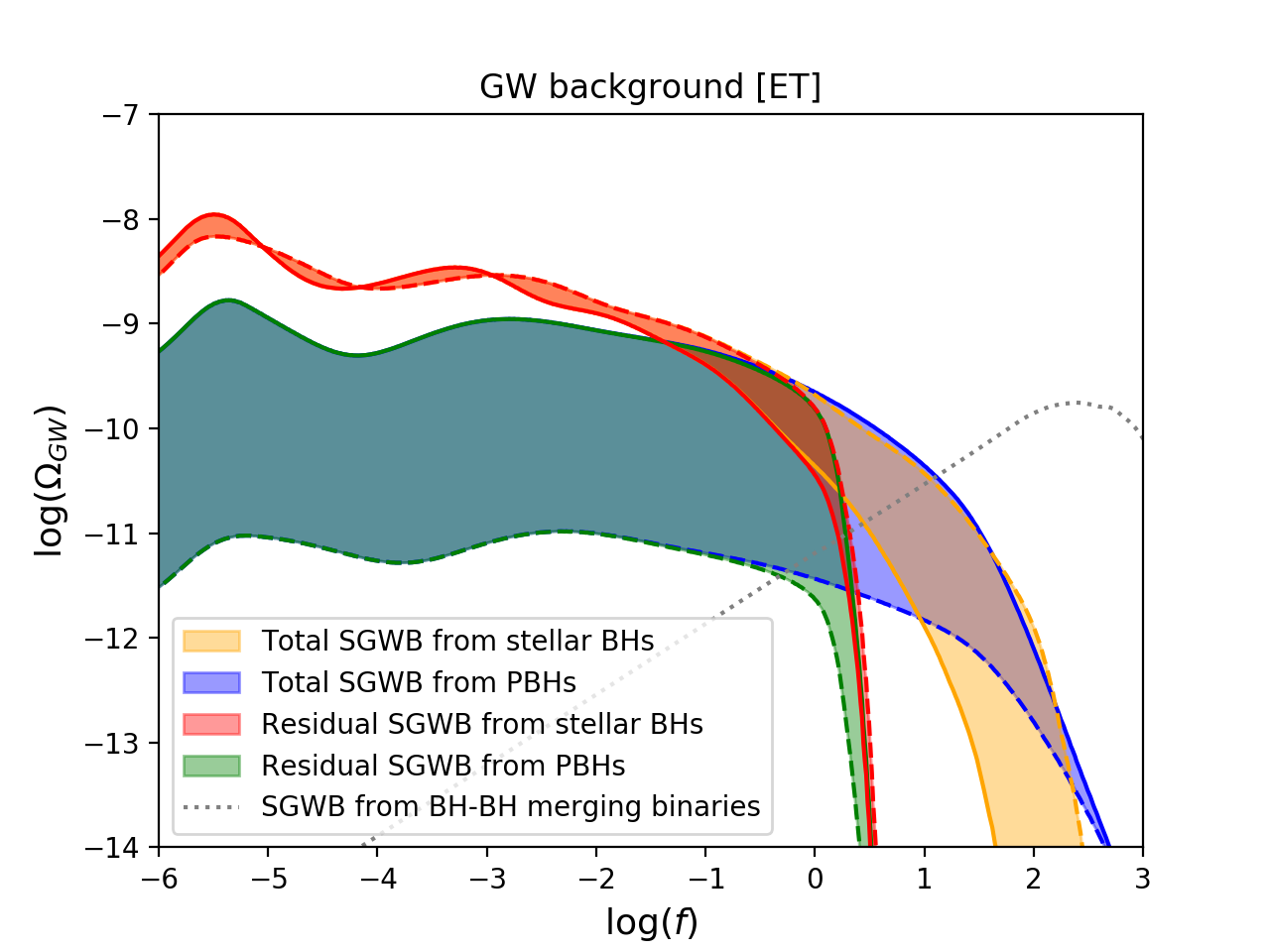}
    \caption{Stochastic GW background seen by ET.
    Yellow shaded area illustrates the background from all stellar compact remnants, blue shaded area from all pBHs, red shaded area from undetected stellar compact remnants, and green shaded area from undetected pBHs. All shaded areas show the effect of varying the pBH-to-DM fraction $f_{\rm pBH}$ from $0.01$ (dashed edge lines) to $1$ (solid edge lines). The dotted line illustrates the background from undetected mergers of isolated BH-BH binaries computed by \citep{boco+19}.}
    \label{fig:SGWB_ET}
\end{figure*}

\begin{figure*}
    \centering
    \includegraphics[width=1.\textwidth]{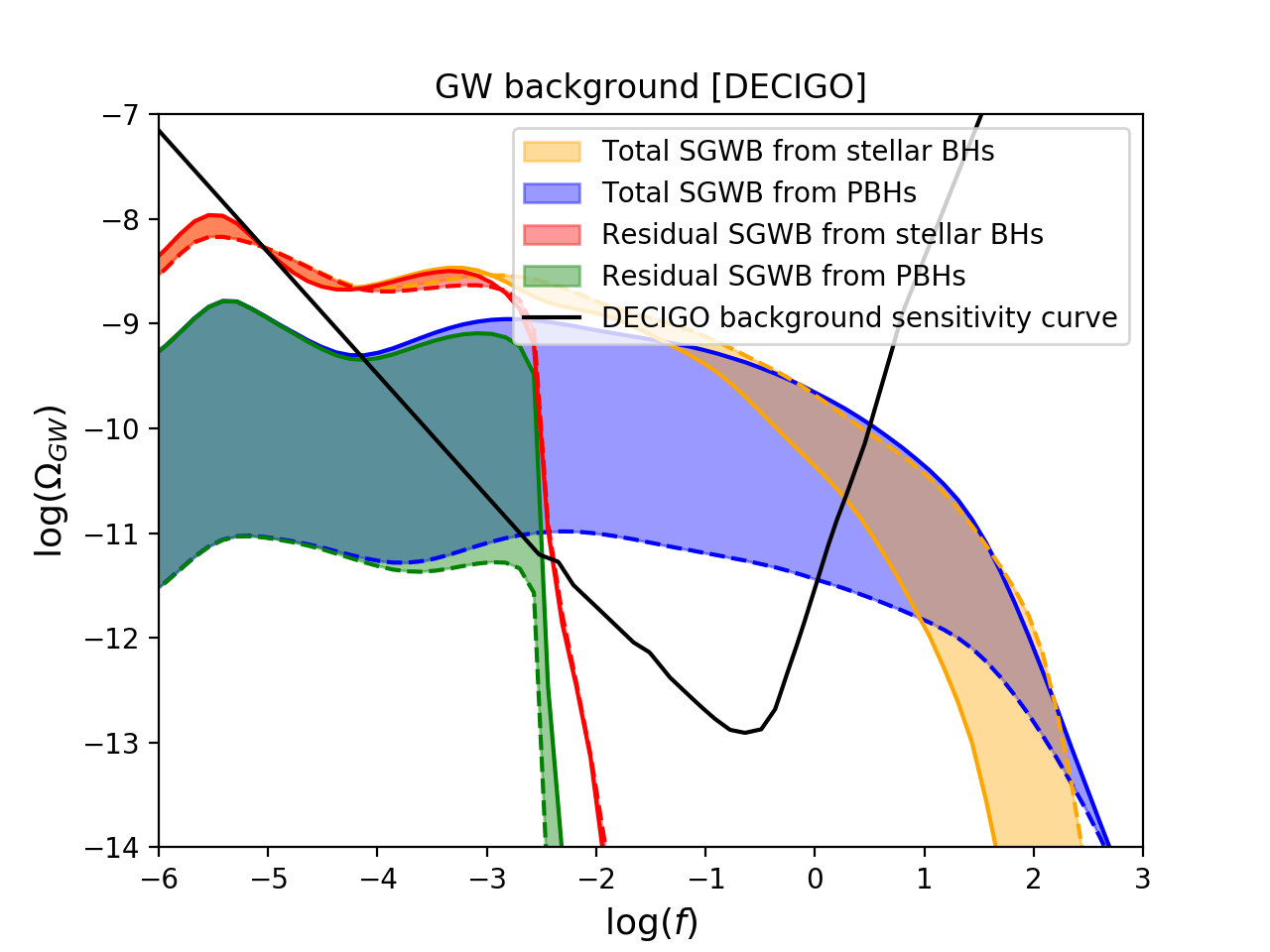}
    \caption{Same as previous figures for DECIGO, whose sensitivity curve for the background is shown as a black solid line.}
    \label{fig:SGWB_DECIGO}
\end{figure*}

\begin{figure*}
    \centering
    \includegraphics[width=1.\textwidth]{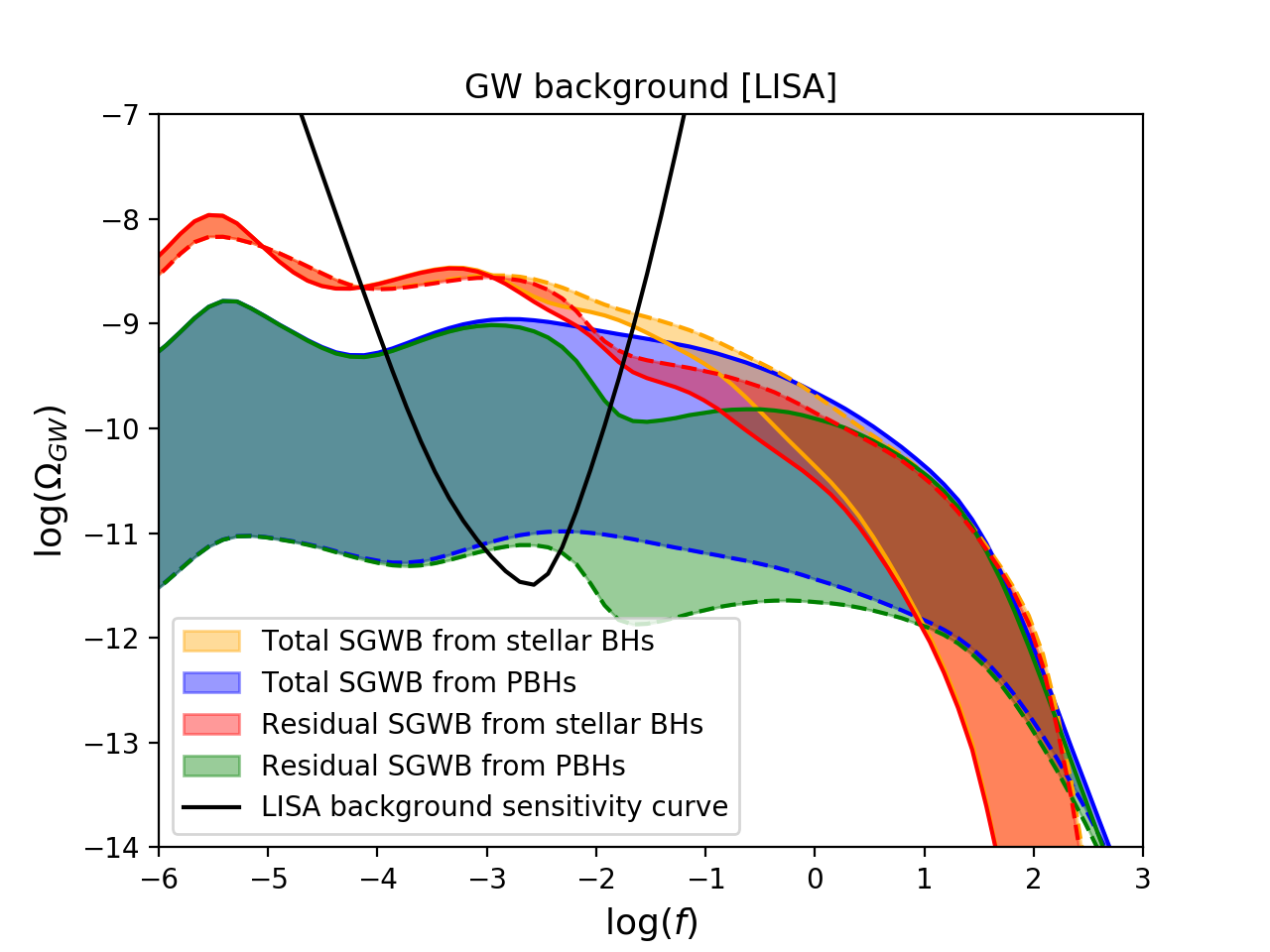}
    \caption{Same as previous figure for LISA, whose sensitivity curve for the background is shown as a black solid line.}
    \label{fig:SGWB_LISA}
\end{figure*}

\begin{figure*}
    \centering
    \includegraphics[width=0.7\textwidth]{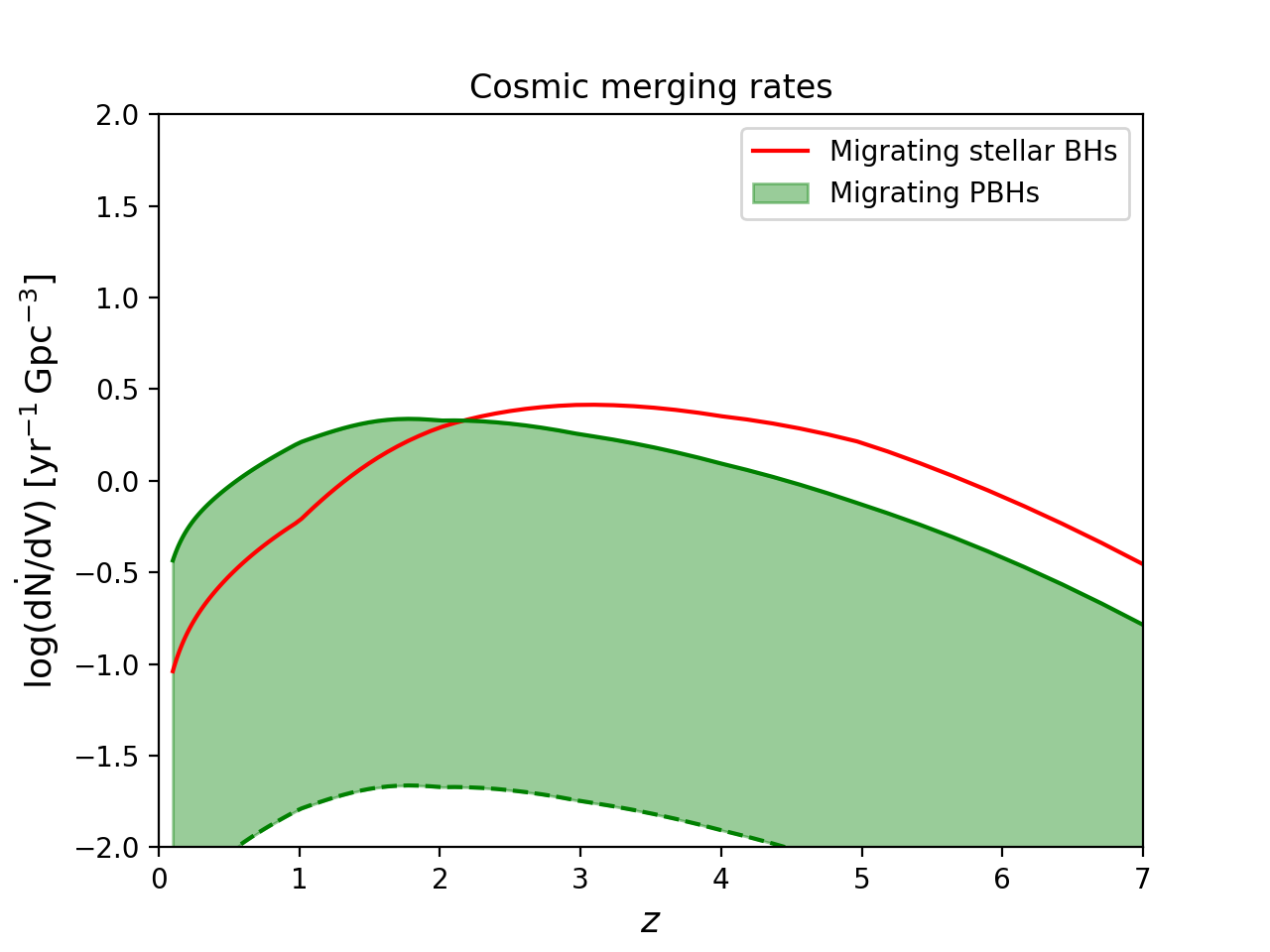}
    \includegraphics[width=0.7\textwidth]{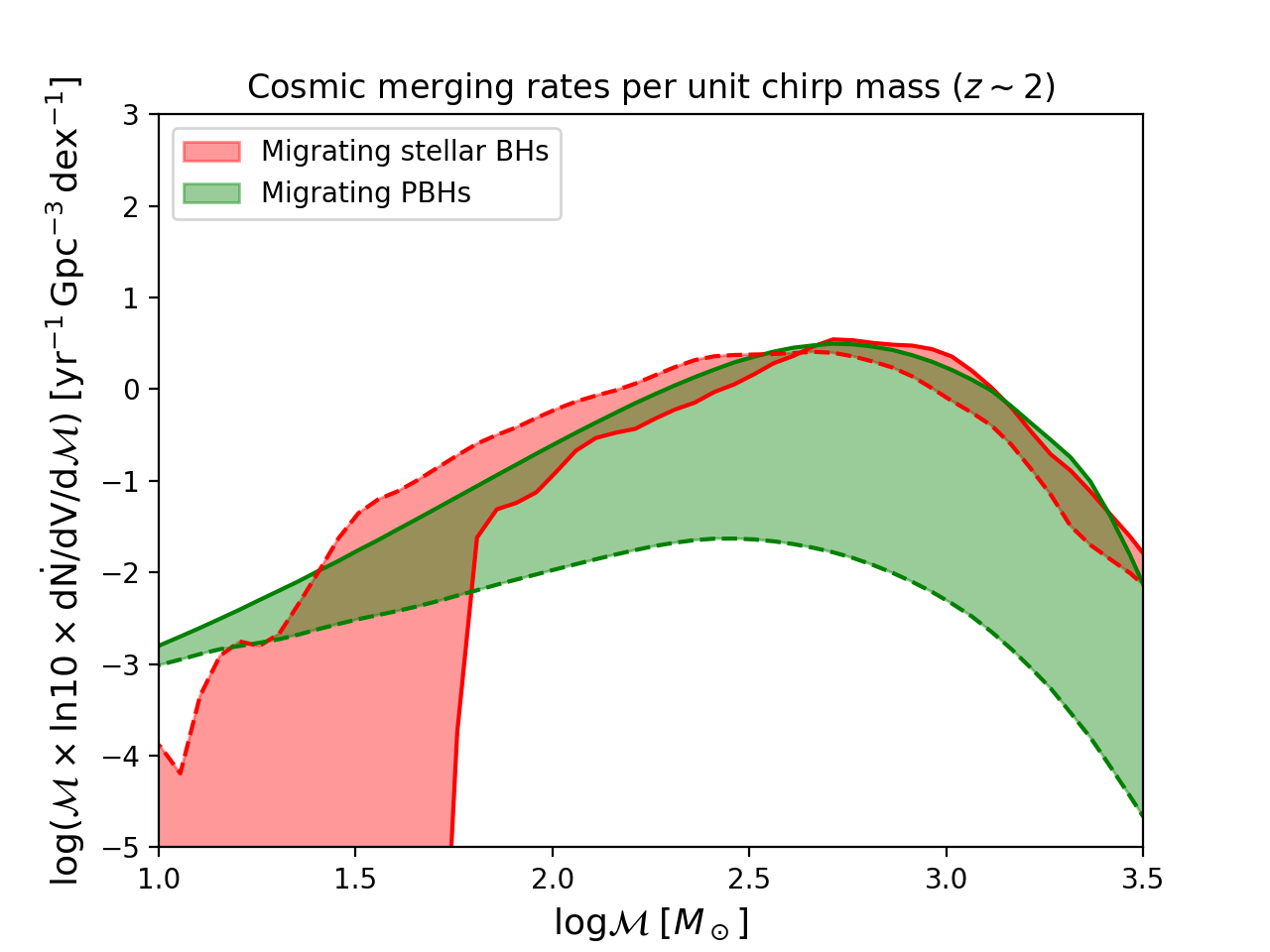}
    \caption{Top panel: same as Fig. \ref{fig:mergingrate} but for $t_{\rm max}\approx 5\times 10^7\,\rm yr$. Bottom panel: same as Fig. \ref{fig:mergingrate_chirp} but for $t_{\rm max}\approx 5\times 10^7\,\rm yr$.}
    \label{fig:mergingrate_stopped}
\end{figure*}

In Figure \ref{fig:mergingrate_stopped} (top panel) we show the merging rates density as a function of redshift. While pBHs mergers are reduced by a factor $\sim 10$, stellar compact remnants mergers are drastically cut by a factor $\sim 10^2$. The stronger impact on stellar compact remnants mergers can be understood since they tend to merge at later times, as seen from Fig. \ref{fig:seed}. Merging rates from stellar compact remnants and pBHs now tend to be comparable for $f_{\rm pBH}\sim 1$, with a prevalence of pBH at $z\lesssim 2$ and of stellar compact remnants at $z\gtrsim 2$.
In Fig. \ref{fig:mergingrate_stopped} (bottom panel) we also present the chirp mass distribution of the merger events at $z\sim 2$. Since the process is stopped at a smaller $t_{\rm max}\approx 5\times 10^7\,\rm yr$, when the central BH has still a mass $M_\bullet\lesssim 10^5\,\rm M_\odot$, the chirp mass distribution cannot extend above $\mathcal{M}_{\bullet\bullet}\sim 3000\,\rm M_\odot$. 

\begin{figure*}
    \centering
    \includegraphics[width=1.\textwidth]{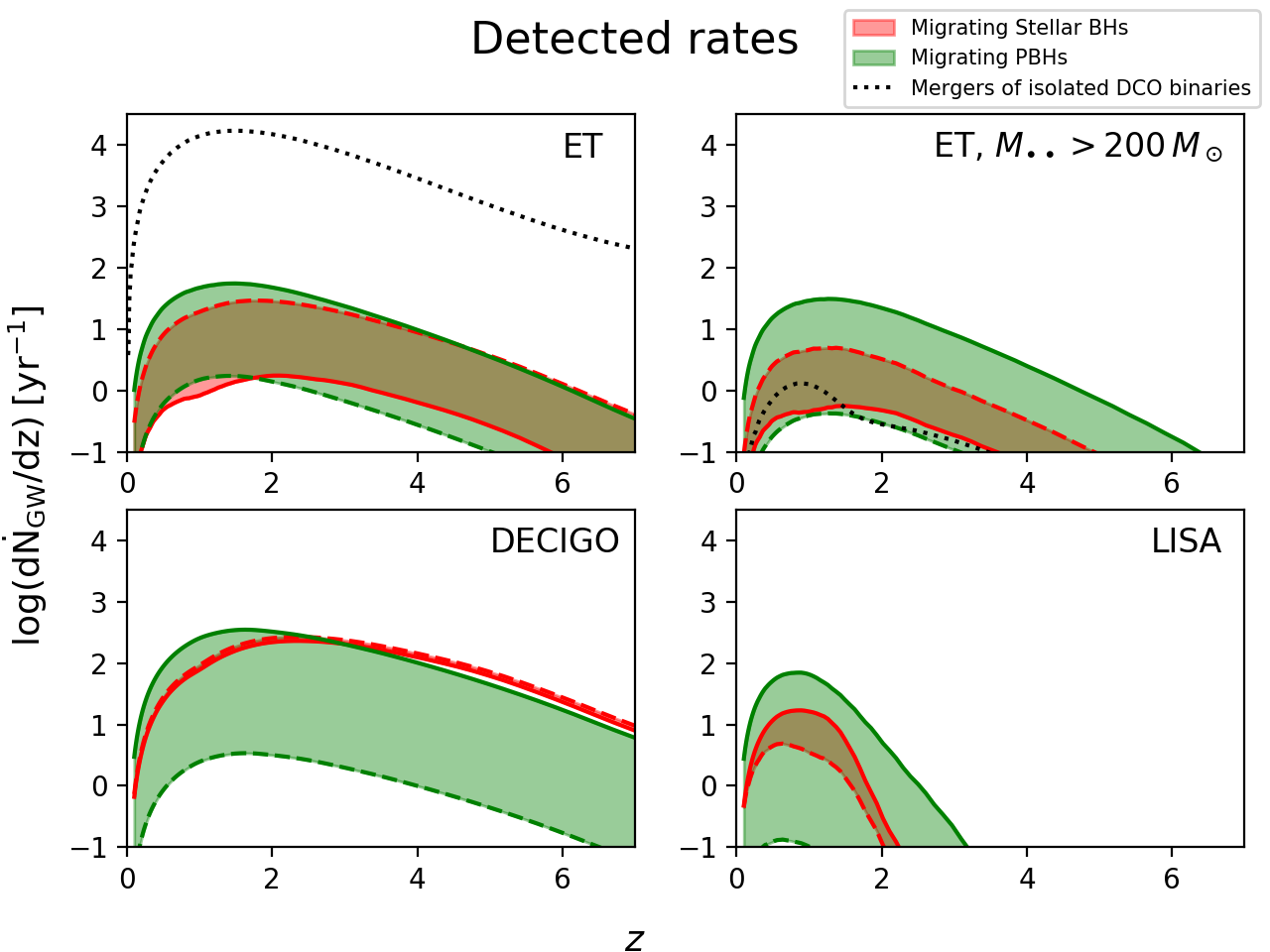}
    \caption{same as Fig. \ref{fig:detected_all} but for $t_{\rm max}\approx 5\times 10^7\,\rm yr$.}
    \label{fig:detected_all_stopped}
\end{figure*}

In Fig. \ref{fig:detected_all_stopped} we show the detection rates for ET, DECIGO and LISA as a function of redshift. The reduction of the detection rates for ET (top left panel) is not severe: a factor $\sim 3$ for stellar compact remnants and almost no reduction for pBHs. This is due to the fact that ET tends to detect low chirp mass events ($\mathcal{M}_{\bullet\bullet}\leq 500\,\rm M_\odot$) occurring during the early stages of the process; its detection rate is therefore only partially affected by a cut in the number of mergers at intermediate and late times. A slightly larger reduction can be seen in the ET detection rates for events with $\mathcal{M}_{\bullet\bullet}>200\,\rm M_\odot$ (top right panel); in particular, detected events from  stellar compact remnants are reduced by a factor $\sim 5$ and from pBHs by a factor $\sim 1.5$.
For DECIGO (bottom left panel), probing both small and intermediate chirp mass regimes, the reduction is stronger with respect to ET: a factor $\sim 30$ for stellar compact remnants and $\sim 3$ for pBHs; however, a significant number of events per year is still detectable.  A dramatic effect can be seen for LISA (bottom right panel); the reduction of the detected events somewhat depends on redshift and on $f_{\rm pBH}$, being of the order of $\sim 10^2$ for stellar compact remnants and $\sim 10$ for pBHs. This is because LISA probes higher chirp masses $\mathcal{M}_{\bullet\bullet}\sim 1000-5000\,\rm M_\odot$ with respect to ET, where merger rates are suppressed (see bottom panel of Fig. \ref{fig:mergingrate_stopped}). 

\begin{figure*}
    \centering
    \includegraphics[width=0.8\textwidth]{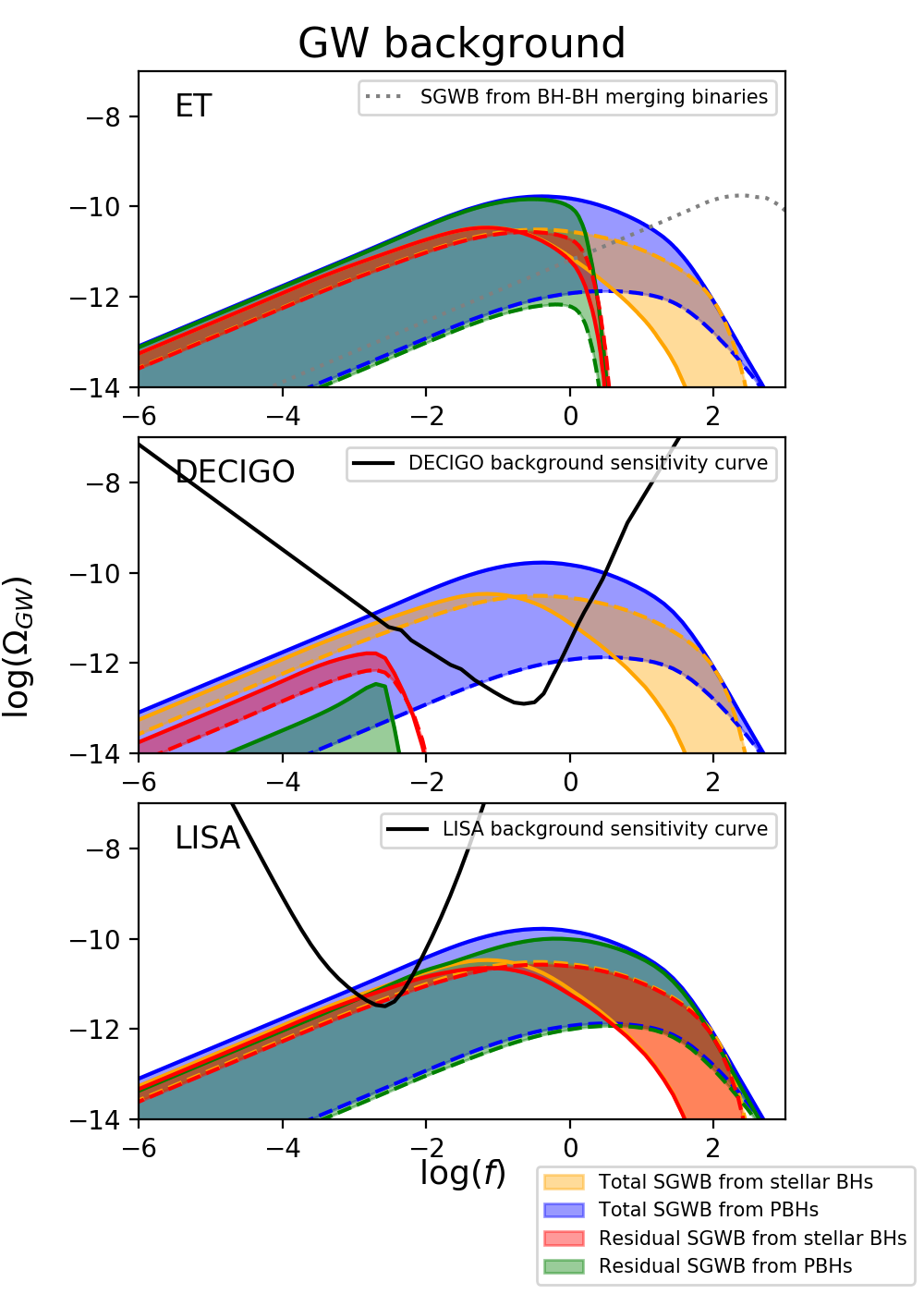}
    \caption{Top panel: same as Fig. \ref{fig:SGWB_ET} but for $t_{\rm max}\approx 5\times 10^7\,\rm yr$. Middle panel: same as Fig. \ref{fig:SGWB_DECIGO} but for $t_{\rm max} \approx 5\times 10^7\,\rm yr$. Bottom panel: same as Fig. \ref{fig:SGWB_LISA} but for $t_{\rm max}\approx 5\times 10^7\,\rm yr$.}
    \label{fig:SGWB_stopped}
\end{figure*}

Finally, in Fig. \ref{fig:SGWB_stopped} we show the predictions for the SGWB generated by all the merging events (orange and blue patches for stellar compact remnants and pBHs, respectively) and by the residual unresolved events for ET (top panel), for DECIGO (middle panel) and for LISA (bottom panel). We notice that there is an overall decrease of the intensity of the SGWB, especially at low frequencies $f\lesssim 0.1\,\rm Hz$, since late time mergers of a massive central BH contributing at those frequencies are cut away.

\end{appendix}

\bibliographystyle{unsrt}  
\bibliography{ref}

\begin{thebibliography}{10}

\bibitem{boco+20}
L.~{Boco}, A.~{Lapi}, and L.~{Danese}.
\newblock {Growth of Supermassive Black Hole Seeds in ETG Star-forming
  Progenitors: Multiple Merging of Stellar Compact Remnants via Gaseous
  Dynamical Friction and Gravitational-wave Emission}.
\newblock {\em The Astrophysical Journal}, 891(1):94, March 2020.

\bibitem{abbott+16}
B.~P. {Abbott}, R.~{Abbott}, T.~D. {Abbott}, et~al.
\newblock {Observation of Gravitational Waves from a Binary Black Hole Merger}.
\newblock {\em Phys. Rev. Lett.}, 116(6):061102, February 2016.

\bibitem{abbott+19a}
B.~P. {Abbott}, R.~{Abbott}, T.~D. {Abbott}, et~al.
\newblock {GWTC-1: A Gravitational-Wave Transient Catalog of Compact Binary
  Mergers Observed by LIGO and Virgo during the First and Second Observing
  Runs}.
\newblock {\em Physical Review X}, 9(3):031040, July 2019.

\bibitem{abbott+20b}
R.~{Abbott}, T.~D. {Abbott}, S.~{Abraham}, et~al.
\newblock {GWTC-2: Compact Binary Coalescences Observed by LIGO and Virgo
  During the First Half of the Third Observing Run}.
\newblock {\em arXiv e-prints}, page arXiv:2010.14527, October 2020.

\bibitem{abbott+19b}
B.~P. {Abbott}, R.~{Abbott}, T.~D. {Abbott}, et~al.
\newblock {Binary Black Hole Population Properties Inferred from the First and
  Second Observing Runs of Advanced LIGO and Advanced Virgo}.
\newblock {\em ApJ Letters}, 882(2):L24, September 2019.

\bibitem{abbott+20a}
R.~{Abbott}, T.~D. {Abbott}, S.~{Abraham}, et~al.
\newblock {GW190521: A Binary Black Hole Merger with a Total Mass of 150
  M$_\odot$}.
\newblock {\em Phys. Rev. Lett.}, 125(10):101102, September 2020.

\bibitem{abbott+20c}
R.~{Abbott}, T.~D. {Abbott}, S.~{Abraham}, et~al.
\newblock {Population Properties of Compact Objects from the Second LIGO-Virgo
  Gravitational-Wave Transient Catalog}.
\newblock {\em arXiv e-prints}, page arXiv:2010.14533, October 2020.

\bibitem{abbott+17a}
B.~P. {Abbott}, R.~{Abbott}, T.~D. {Abbott}, et~al.
\newblock {A gravitational-wave standard siren measurement of the Hubble
  constant}.
\newblock {\em Nature}, 551(7678):85--88, November 2017.

\bibitem{abbott+21}
B.~P. {Abbott}, R.~{Abbott}, T.~D. {Abbott}, et~al.
\newblock {A Gravitational-wave Measurement of the Hubble Constant Following
  the Second Observing Run of Advanced LIGO and Virgo}.
\newblock {\em The Astrophysical Journal}, 909(2):218, March 2021.

\bibitem{hotokezaka+19}
K.~{Hotokezaka}, E.~{Nakar}, O.~{Gottlieb}, et~al.
\newblock {A Hubble constant measurement from superluminal motion of the jet in
  GW170817}.
\newblock {\em Nature Astronomy}, 3:940--944, July 2019.

\bibitem{soares-santos+19}
M.~{Soares-Santos}, A.~{Palmese}, W.~{Hartley}, et~al.
\newblock {First Measurement of the Hubble Constant from a Dark Standard Siren
  using the Dark Energy Survey Galaxies and the LIGO/Virgo Binary-Black-hole
  Merger GW170814}.
\newblock {\em ApJ Letters}, 876(1):L7, May 2019.

\bibitem{mukherjee+21}
Suvodip {Mukherjee}, Guilhem {Lavaux}, Fran{\c{c}}ois~R. {Bouchet}, et~al.
\newblock {Velocity correction for Hubble constant measurements from standard
  sirens}.
\newblock {\em A\&A}, 646:A65, February 2021.

\bibitem{abbott+17b}
B.~P. {Abbott}, R.~{Abbott}, T.~D. {Abbott}, et~al.
\newblock {Gravitational Waves and Gamma-Rays from a Binary Neutron Star
  Merger: GW170817 and GRB 170817A}.
\newblock {\em ApJ Letters}, 848(2):L13, October 2017.

\bibitem{creminelli+17}
Paolo {Creminelli} and Filippo {Vernizzi}.
\newblock {Dark Energy after GW170817 and GRB170817A}.
\newblock {\em Phys. Rev. Lett.}, 119(25):251302, December 2017.

\bibitem{belczynski+16}
Krzysztof {Belczynski}, Daniel~E. {Holz}, Tomasz {Bulik}, and Richard
  {O'Shaughnessy}.
\newblock {The first gravitational-wave source from the isolated evolution of
  two stars in the 40-100 solar mass range}.
\newblock {\em Nature}, 534(7608):512--515, June 2016.

\bibitem{lamberts+16}
A.~{Lamberts}, S.~{Garrison-Kimmel}, D.~R. {Clausen}, and P.~F. {Hopkins}.
\newblock {When and where did GW150914 form?}
\newblock {\em Monthly Notices of the Royal Astronomical Society},
  463(1):L31--L35, November 2016.

\bibitem{cao+18}
Liang {Cao}, Youjun {Lu}, and Yuetong {Zhao}.
\newblock {Host galaxy properties of mergers of stellar binary black holes and
  their implications for advanced LIGO gravitational wave sources}.
\newblock {\em Monthly Notices of the Royal Astronomical Society},
  474(4):4997--5007, March 2018.

\bibitem{elbert+18}
Oliver~D. {Elbert}, James~S. {Bullock}, and Manoj {Kaplinghat}.
\newblock {Counting black holes: The cosmic stellar remnant population and
  implications for LIGO}.
\newblock {\em Monthly Notices of the Royal Astronomical Society},
  473(1):1186--1194, January 2018.

\bibitem{li+18}
Shun-Sheng {Li}, Shude {Mao}, Yuetong {Zhao}, and Youjun {Lu}.
\newblock {Gravitational lensing of gravitational waves: a statistical
  perspective}.
\newblock {\em Monthly Notices of the Royal Astronomical Society},
  476(2):2220--2229, May 2018.

\bibitem{mapelli+18}
Michela {Mapelli} and Nicola {Giacobbo}.
\newblock {The cosmic merger rate of neutron stars and black holes}.
\newblock {\em Monthly Notices of the Royal Astronomical Society},
  479(4):4391--4398, October 2018.

\bibitem{artale+19}
M.~Celeste {Artale}, Michela {Mapelli}, Nicola {Giacobbo}, et~al.
\newblock {Host galaxies of merging compact objects: mass, star formation rate,
  metallicity, and colours}.
\newblock {\em Monthly Notices of the Royal Astronomical Society},
  487(2):1675--1688, August 2019.

\bibitem{boco+19}
L.~{Boco}, A.~{Lapi}, S.~{Goswami}, et~al.
\newblock {Merging Rates of Compact Binaries in Galaxies: Perspectives for
  Gravitational Wave Detections}.
\newblock {\em The Astrophyscal Journal}, 7 2019.

\bibitem{neijssel+19}
Coenraad~J. {Neijssel}, Alejandro {Vigna-G{\'o}mez}, Simon {Stevenson}, et~al.
\newblock {The effect of the metallicity-specific star formation history on
  double compact object mergers}.
\newblock {\em Monthly Notices of the Royal Astronomical Society},
  490(3):3740--3759, December 2019.

\bibitem{boco+21}
L.~{Boco}, A.~{Lapi}, M.~{Chruslinska}, et~al.
\newblock {Evolution of Galaxy Star Formation and Metallicity: Impact on Double
  Compact Object Mergers}.
\newblock {\em The Astrphysical journal}, 907(2):110, February 2021.

\bibitem{santoliquido+21}
Filippo {Santoliquido}, Michela {Mapelli}, Nicola {Giacobbo}, Yann
  {Bouffanais}, and M.~Celeste {Artale}.
\newblock {The cosmic merger rate density of compact objects: impact of star
  formation, metallicity, initial mass function, and binary evolution}.
\newblock {\em Monthly Notices of the Royal Astronomical Society},
  502(4):4877--4889, April 2021.

\bibitem{clesse+17}
S{\'e}bastien {Clesse} and Juan {Garc{\'\i}a-Bellido}.
\newblock {Detecting the gravitational wave background from primordial black
  hole dark matter}.
\newblock {\em Physics of the Dark Universe}, 18:105--114, December 2017.

\bibitem{raidal+17}
Martti {Raidal}, Ville {Vaskonen}, and Hardi {Veerm{\"a}e}.
\newblock {Gravitational waves from primordial black hole mergers}.
\newblock {\em JCAP}, 2017(9):037, September 2017.

\bibitem{scelfo+18}
Giulio {Scelfo}, Nicola {Bellomo}, Alvise {Raccanelli}, Sabino {Matarrese}, and
  Licia {Verde}.
\newblock {GW{\texttimes}LSS: chasing the progenitors of merging binary black
  holes}.
\newblock {\em JCAP}, 2018(9):039, September 2018.

\bibitem{kimura+21}
Rampei {Kimura}, Teruaki {Suyama}, Masahide {Yamaguchi}, and Ying-li {Zhang}.
\newblock {Reconstruction of primordial power spectrum of curvature
  perturbation from the merger rate of primordial black hole binaries}.
\newblock {\em JCAP}, 2021(4):031, April 2021.

\bibitem{wang+16}
Sai Wang, Yi-Fan Wang, Qing-Guo Huang, and Tjonnie G.~F. Li.
\newblock {Constraints on the Primordial Black Hole Abundance from the First
  Advanced LIGO Observation Run Using the Stochastic Gravitational-Wave
  Background}.
\newblock {\em Phys. Rev. Lett.}, 120(19):191102, 2018.

\bibitem{wang+19}
Sai Wang, Takahiro Terada, and Kazunori Kohri.
\newblock {Prospective constraints on the primordial black hole abundance from
  the stochastic gravitational-wave backgrounds produced by coalescing events
  and curvature perturbations}.
\newblock {\em Physical Review D}, 99(10):103531, 2019.
\newblock [Erratum: Phys.Rev.D 101, 069901 (2020)].

\bibitem{scoville+14}
N.~{Scoville}, H.~{Aussel}, K.~{Sheth}, et~al.
\newblock {The Evolution of Interstellar Medium Mass Probed by Dust Emission:
  ALMA Observations at z = 0.3-2}.
\newblock {\em The Astrophysical Journal}, 783(2):84, March 2014.

\bibitem{scoville+16}
N.~{Scoville}, K.~{Sheth}, H.~{Aussel}, et~al.
\newblock {ISM Masses and the Star formation Law at Z = 1 to 6: ALMA
  Observations of Dust Continuum in 145 Galaxies in the COSMOS Survey Field}.
\newblock {\em The Astrophysical Journal}, 820(2):83, April 2016.

\bibitem{ikarashi+15}
Soh {Ikarashi}, R.~J. {Ivison}, Karina~I. {Caputi}, et~al.
\newblock {Compact Starbursts in z {\ensuremath{\sim}} 3-6 Submillimeter
  Galaxies Revealed by ALMA}.
\newblock {\em The Astrophysical Journal}, 810(2):133, September 2015.

\bibitem{simpson+15}
J.~M. {Simpson}, Ian {Smail}, A.~M. {Swinbank}, et~al.
\newblock {The SCUBA-2 Cosmology Legacy Survey: ALMA Resolves the Bright-end of
  the Sub-millimeter Number Counts}.
\newblock {\em The Astrophysical Journal}, 807(2):128, July 2015.

\bibitem{barro+16}
Guillermo {Barro}, Sandra~M. {Faber}, Avishai {Dekel}, et~al.
\newblock {Caught in the Act: Gas and Stellar Velocity Dispersions in a Fast
  Quenching Compact Star-Forming Galaxy at
  z\raisebox{-0.5ex}\textasciitilde1.7}.
\newblock {\em The Astrophysical Journal}, 820(2):120, April 2016.

\bibitem{spilker+16}
J.~S. {Spilker}, D.~P. {Marrone}, M.~{Aravena}, et~al.
\newblock {ALMA Imaging and Gravitational Lens Models of South Pole
  Telescope{\textemdash}Selected Dusty, Star-Forming Galaxies at High
  Redshifts}.
\newblock {\em The Astrophysical Journal}, 826(2):112, August 2016.

\bibitem{tadaki+17a}
Ken-ichi {Tadaki}, Reinhard {Genzel}, Tadayuki {Kodama}, et~al.
\newblock {Bulge-forming Galaxies with an Extended Rotating Disk at z
  \raisebox{-0.5ex}\textasciitilde 2}.
\newblock {\em The Astrophysical Journal}, 834(2):135, January 2017.

\bibitem{tadaki+17b}
Ken-ichi {Tadaki}, Tadayuki {Kodama}, Erica~J. {Nelson}, et~al.
\newblock {Rotating Starburst Cores in Massive Galaxies at z = 2.5}.
\newblock {\em ApJ Letters}, 841(2):L25, June 2017.

\bibitem{tadaki+18}
K.~{Tadaki}, D.~{Iono}, M.~S. {Yun}, et~al.
\newblock {The gravitationally unstable gas disk of a starburst galaxy 12
  billion years ago}.
\newblock {\em Nature}, 560(7720):613--616, August 2018.

\bibitem{talia+18}
M.~{Talia}, F.~{Pozzi}, L.~{Vallini}, et~al.
\newblock {ALMA view of a massive spheroid progenitor: a compact rotating core
  of molecular gas in an AGN host at z = 2.226}.
\newblock {\em Monthly Notices of the Royal Astronomical Society},
  476(3):3956--3963, May 2018.

\bibitem{lang+19}
Philipp {Lang}, E.~{Schinnerer}, Ian {Smail}, et~al.
\newblock {Revealing the Stellar Mass and Dust Distributions of Submillimeter
  Galaxies at Redshift 2}.
\newblock {\em The Astrophysical Journal}, 879(1):54, July 2019.

\bibitem{smail+21}
Ian {Smail}, U.~{Dudzevi{\v{c}}i{\={u}}t{\.{e}}}, S.~M. {Stach}, et~al.
\newblock {An ALMA survey of the S2CLS UDS field: optically invisible
  submillimetre galaxies}.
\newblock {\em Monthly Notices of the Royal Astronomical Society},
  502(3):3426--3435, April 2021.

\bibitem{planck2020}
{Planck Collaboration}.
\newblock {Planck 2018 results. VI. Cosmological parameters}.
\newblock {\em A\&A}, 641:A6, September 2020.

\bibitem{chandrasekhar+43}
S.~{Chandrasekhar}.
\newblock {Dynamical Friction. I. General Considerations: the Coefficient of
  Dynamical Friction.}
\newblock {\em The Astrophysical Journal}, 97:255, March 1943.

\bibitem{binney+87}
James {Binney} and Scott {Tremaine}.
\newblock {\em {Galactic dynamics}}.
\newblock 1987.

\bibitem{lacey+93}
Cedric {Lacey} and Shaun {Cole}.
\newblock {Merger rates in hierarchical models of galaxy formation}.
\newblock {\em Monthly Notices of the Royal Astronomical Society},
  262(3):627--649, June 1993.

\bibitem{hashimoto+03}
Yoshikazu {Hashimoto}, Yoko {Funato}, and Junichiro {Makino}.
\newblock {To Circularize or Not To Circularize?-Orbital Evolution of Satellite
  Galaxies}.
\newblock {\em The Astrophysical Journal}, 582(1):196--201, January 2003.

\bibitem{fujii+06}
Michiko {Fujii}, Yoko {Funato}, and Junichiro {Makino}.
\newblock {Dynamical Friction on Satellite Galaxies}.
\newblock {\em PASJ}, 58:743--752, August 2006.

\bibitem{boylan-kolchin+08}
Michael {Boylan-Kolchin}, Chung-Pei {Ma}, and Eliot {Quataert}.
\newblock {Dynamical friction and galaxy merging time-scales}.
\newblock {\em Monthly Notices of the Royal Astronomical Society},
  383(1):93--101, January 2008.

\bibitem{jiang+08}
C.~Y. {Jiang}, Y.~P. {Jing}, A.~{Faltenbacher}, W.~P. {Lin}, and Cheng {Li}.
\newblock {A Fitting Formula for the Merger Timescale of Galaxies in
  Hierarchical Clustering}.
\newblock {\em The Astrophysical Journal}, 675(2):1095--1105, March 2008.

\bibitem{begelman+80}
M.~C. {Begelman}, R.~D. {Blandford}, and M.~J. {Rees}.
\newblock {Massive black hole binaries in active galactic nuclei}.
\newblock {\em Nature}, 287(5780):307--309, September 1980.

\bibitem{mayer+07}
L.~{Mayer}, S.~{Kazantzidis}, P.~{Madau}, et~al.
\newblock {Rapid Formation of Supermassive Black Hole Binaries in Galaxy
  Mergers with Gas}.
\newblock {\em Science}, 316(5833):1874, June 2007.

\bibitem{barausse+12}
Enrico {Barausse}.
\newblock {The evolution of massive black holes and their spins in their
  galactic hosts}.
\newblock {\em Monthly Notices of the Royal Astronomical Society},
  423(3):2533--2557, July 2012.

\bibitem{chapon+13}
Damien {Chapon}, Lucio {Mayer}, and Romain {Teyssier}.
\newblock {Hydrodynamics of galaxy mergers with supermassive black holes: is
  there a last parsec problem?}
\newblock {\em Monthly Notices of the Royal Astronomical Society},
  429(4):3114--3122, March 2013.

\bibitem{antonini+15}
Fabio {Antonini}, Enrico {Barausse}, and Joseph {Silk}.
\newblock {The Imprint of Massive Black Hole Mergers on the Correlation between
  Nuclear Star Clusters and Their Host Galaxies}.
\newblock {\em ApJ Letters}, 806(1):L8, June 2015.

\bibitem{tamburello+17}
Valentina {Tamburello}, Pedro~R. {Capelo}, Lucio {Mayer}, Jillian~M.
  {Bellovary}, and James~W. {Wadsley}.
\newblock {Supermassive black hole pairs in clumpy galaxies at high redshift:
  delayed binary formation and concurrent mass growth}.
\newblock {\em Monthly Notices of the Royal Astronomical Society},
  464(3):2952--2962, January 2017.

\bibitem{dayal+19}
Pratika {Dayal}, Elena~M. {Rossi}, Banafsheh {Shiralilou}, et~al.
\newblock {The hierarchical assembly of galaxies and black holes in the first
  billion years: predictions for the era of gravitational wave astronomy}.
\newblock {\em Monthly Notices of the Royal Astronomical Society},
  486(2):2336--2350, June 2019.

\bibitem{katz+20}
Michael~L. {Katz}, Luke~Zoltan {Kelley}, Fani {Dosopoulou}, Samantha {Berry},
  Laura {Blecha}, and Shane~L. {Larson}.
\newblock {Probing massive black hole binary populations with LISA}.
\newblock {\em Monthly Notices of the Royal Astronomical Society},
  491(2):2301--2317, January 2020.

\bibitem{dokuchaev+64}
V.~P. {Dokuchaev}.
\newblock {Emission of Magnetoacoustic Waves in the Motion of Stars in Cosmic
  Space.}
\newblock {\em SvA}, 8:23, August 1964.

\bibitem{ruderman+17}
M.~A. {Ruderman} and E.~A. {Spiegel}.
\newblock {Galactic Wakes}.
\newblock {\em The Astrophysical Journal}, 165:1, April 1971.

\bibitem{bisnovatyi-kogan+79}
G.~S. {Bisnovatyi-Kogan}, Ya.~M. {Kazhdan}, A.~A. {Klypin}, A.~E. {Lutskii},
  and N.~I. {Shakura}.
\newblock {Accretion onto a rapidly moving gravitating center}.
\newblock {\em SvA}, 23:201--205, April 1979.

\bibitem{rephaeli+80}
Y.~{Rephaeli} and E.~E. {Salpeter}.
\newblock {Flow past a massive object and the gravitational drag}.
\newblock {\em The Astrophysical Journal}, 240:20--24, August 1980.

\bibitem{ostriker+99}
Eve~C. {Ostriker}.
\newblock {Dynamical Friction in a Gaseous Medium}.
\newblock {\em The Astrophysical Journal}, 513(1):252--258, March 1999.

\bibitem{sanchez-salcedo+01}
F.~J. {S{\'a}nchez-Salcedo} and A.~{Brandenburg}.
\newblock {Dynamical friction of bodies orbiting in a gaseous sphere}.
\newblock {\em Monthly Notices of the Royal Astronomical Society},
  322(1):67--78, March 2001.

\bibitem{escala+04}
Andr{\'e}s {Escala}, Richard~B. {Larson}, Paolo~S. {Coppi}, and Diego
  {Mardones}.
\newblock {The Role of Gas in the Merging of Massive Black Holes in Galactic
  Nuclei. I. Black Hole Merging in a Spherical Gas Cloud}.
\newblock {\em The Astrophysical Journal}, 607(2):765--777, June 2004.

\bibitem{tanaka+09}
Takamitsu {Tanaka} and Zolt{\'a}n {Haiman}.
\newblock {The Assembly of Supermassive Black Holes at High Redshifts}.
\newblock {\em The Astrophysical Journal}, 696(2):1798--1822, May 2009.

\bibitem{tagawa+16}
H.~{Tagawa}, M.~{Umemura}, and N.~{Gouda}.
\newblock {Mergers of accreting stellar-mass black holes}.
\newblock {\em Monthly Notices of the Royal Astronomical Society},
  462(4):3812--3822, November 2016.

\bibitem{prugniel+97}
P.~{Prugniel} and F.~{Simien}.
\newblock {The fundamental plane of early-type galaxies: non-homology of the
  spatial structure.}
\newblock {\em A\&A}, 321:111--122, May 1997.

\bibitem{speagle+14}
J.~S. {Speagle}, C.~L. {Steinhardt}, P.~L. {Capak}, and J.~D. {Silverman}.
\newblock {A Highly Consistent Framework for the Evolution of the Star-Forming
  ``Main Sequence'' from z \raisebox{-0.5ex}\textasciitilde 0-6}.
\newblock {\em The Astrophysical Journal Supplement Series}, 214(2):15, October
  2014.

\bibitem{lapi+17a}
A.~{Lapi}, C.~{Mancuso}, A.~{Bressan}, and L.~{Danese}.
\newblock {Stellar Mass Function of Active and Quiescent Galaxies via the
  Continuity Equation}.
\newblock {\em The Astrophysical Journal}, 847(1):13, September 2017.

\bibitem{moster+13}
Benjamin~P. {Moster}, Thorsten {Naab}, and Simon D.~M. {White}.
\newblock {Galactic star formation and accretion histories from matching
  galaxies to dark matter haloes}.
\newblock {\em Monthly Notices of the Royal Astronomical Society},
  428(4):3121--3138, February 2013.

\bibitem{aversa+15}
R.~{Aversa}, A.~{Lapi}, G.~{de Zotti}, F.~{Shankar}, and L.~{Danese}.
\newblock {Black Hole and Galaxy Coevolution from Continuity Equation and
  Abundance Matching}.
\newblock {\em The Astrophysical Journal}, 810(1):74, September 2015.

\bibitem{shi+17}
J.~{Shi}, A.~{Lapi}, C.~{Mancuso}, H.~{Wang}, and L.~{Danese}.
\newblock {Angular Momentum of Early- and Late-type Galaxies: Nature or
  Nurture?}
\newblock {\em The Astrophysical Journal}, 843(2):105, July 2017.

\bibitem{behroozi+19}
Peter {Behroozi}, Risa~H. {Wechsler}, Andrew~P. {Hearin}, and Charlie {Conroy}.
\newblock {UNIVERSEMACHINE: The correlation between galaxy growth and dark
  matter halo assembly from z = 0-10}.
\newblock {\em Monthly Notices of the Royal Astronomical Society},
  488(3):3143--3194, September 2019.

\bibitem{belczynski+08}
Krzysztof {Belczynski}, Vassiliki {Kalogera}, Frederic~A. {Rasio}, et~al.
\newblock {Compact Object Modeling with the StarTrack Population Synthesis
  Code}.
\newblock {\em The Astrophysical Journal Supplement Series}, 174(1):223--260,
  January 2008.

\bibitem{dominik+12}
Michal {Dominik}, Krzysztof {Belczynski}, Christopher {Fryer}, et~al.
\newblock {Double Compact Objects. I. The Significance of the Common Envelope
  on Merger Rates}.
\newblock {\em The Astrophysical Journal}, 759(1):52, November 2012.

\bibitem{belczynski+10}
Krzysztof {Belczynski}, Michal {Dominik}, Tomasz {Bulik}, Richard
  {O'Shaughnessy}, Chris {Fryer}, and Daniel~E. {Holz}.
\newblock {The Effect of Metallicity on the Detection Prospects for
  Gravitational Waves}.
\newblock {\em The Astrophysical Journal Letters}, 715(2):L138--L141, June
  2010.

\bibitem{mapelli+21}
Michela {Mapelli}, Marco {Dall'Amico}, Yann {Bouffanais}, et~al.
\newblock {Hierarchical black hole mergers in young, globular and nuclear star
  clusters: the effect of metallicity, spin and cluster properties}.
\newblock {\em Monthly Notices of the Royal Astronomical Society},
  505(1):339--358, July 2021.

\bibitem{hobbs+05}
G.~{Hobbs}, D.~R. {Lorimer}, A.~G. {Lyne}, and M.~{Kramer}.
\newblock {A statistical study of 233 pulsar proper motions}.
\newblock {\em Monthly Notices of the Royal Astronomical Society},
  360(3):974--992, July 2005.

\bibitem{scelfo+20}
Giulio {Scelfo}, Lumen {Boco}, Andrea {Lapi}, and Matteo {Viel}.
\newblock {Exploring galaxies-gravitational waves cross-correlations as an
  astrophysical probe}.
\newblock {\em JCAP}, 2020(10):045, October 2020.

\bibitem{capurri+21}
Giulia {Capurri}, Andrea {Lapi}, Carlo {Baccigalupi}, et~al.
\newblock {Intensity and anisotropies of the stochastic Gravitational Wave
  background from merging compact binaries in galaxies}.
\newblock {\em arXiv e-prints}, page arXiv:2103.12037, March 2021.

\bibitem{sasaki+18}
Misao {Sasaki}, Teruaki {Suyama}, Takahiro {Tanaka}, and Shuichiro {Yokoyama}.
\newblock {Primordial black holes{\textemdash}perspectives in gravitational
  wave astronomy}.
\newblock {\em Classical and Quantum Gravity}, 35(6):063001, March 2018.

\bibitem{carr+10}
B.~J. {Carr}, Kazunori {Kohri}, Yuuiti {Sendouda}, and Jun'Ichi {Yokoyama}.
\newblock {New cosmological constraints on primordial black holes}.
\newblock {\em Physical Review D}, 81(10):104019, May 2010.

\bibitem{carr+17}
Bernard {Carr}, Martti {Raidal}, Tommi {Tenkanen}, Ville {Vaskonen}, and Hardi
  {Veerm{\"a}e}.
\newblock {Primordial black hole constraints for extended mass functions}.
\newblock {\em Physical Review D}, 96(2):023514, July 2017.

\bibitem{carr+21}
Bernard {Carr}, Kazunori {Kohri}, Yuuiti {Sendouda}, and Jun'ichi {Yokoyama}.
\newblock {Constraints on Primordial Black Holes}.
\newblock {\em arXiv e-prints}, page arXiv:2002.12778, February 2020.

\bibitem{davis+11}
Shane~W. {Davis} and Ari {Laor}.
\newblock {The Radiative Efficiency of Accretion Flows in Individual Active
  Galactic Nuclei}.
\newblock {\em The Astrophysical Journal}, 728(2):98, February 2011.

\bibitem{raimundo+12}
S.~I. {Raimundo}, A.~C. {Fabian}, R.~V. {Vasudevan}, P.~{Gandhi}, and Jianfeng
  {Wu}.
\newblock {Can we measure the accretion efficiency of active galactic nuclei?}
\newblock {\em Monthly Notices of the Royal Astronomical Society},
  419(3):2529--2544, January 2012.

\bibitem{wu+13}
Shumei {Wu}, Youjun {Lu}, Fupeng {Zhang}, and Ye~{Lu}.
\newblock {Radiative efficiency of disc accretion in individual SDSS QSOs}.
\newblock {\em Monthly Notices of the Royal Astronomical Society},
  436(4):3271--3285, December 2013.

\bibitem{mancuso+16a}
C.~{Mancuso}, A.~{Lapi}, J.~{Shi}, et~al.
\newblock {The Quest for Dusty Star-forming Galaxies at High Redshift z
  {\ensuremath{\gtrsim}} 4}.
\newblock {\em The Astrophysical Journal}, 823(2):128, June 2016.

\bibitem{mancuso+16b}
C.~{Mancuso}, A.~{Lapi}, J.~{Shi}, et~al.
\newblock {The Main Sequences of Star-forming Galaxies and Active Galactic
  Nuclei at High Redshift}.
\newblock {\em The Astrophysical Journal}, 833(2):152, December 2016.

\bibitem{mancuso+17}
C.~{Mancuso}, A.~{Lapi}, I.~{Prandoni}, et~al.
\newblock {Galaxy Evolution in the Radio Band: The Role of Star-forming
  Galaxies and Active Galactic Nuclei}.
\newblock {\em The Astrophysical Journal}, 842(2):95, June 2017.

\bibitem{lapi+17b}
A.~{Lapi}, C.~{Mancuso}, A.~{Celotti}, and L.~{Danese}.
\newblock {Galaxy Evolution at High Redshift: Obscured Star Formation, GRB
  Rates, Cosmic Reionization, and Missing Satellites}.
\newblock {\em The Astrophysical Journal}, 835(1):37, January 2017.

\bibitem{peters+64}
P.~C. {Peters}.
\newblock {Gravitational Radiation and the Motion of Two Point Masses}.
\newblock {\em Physical Review}, 136(4B):1224--1232, November 1964.

\bibitem{zhu+11}
Xing-Jiang {Zhu}, E.~{Howell}, T.~{Regimbau}, D.~{Blair}, and Zong-Hong {Zhu}.
\newblock {Stochastic Gravitational Wave Background from Coalescing Binary
  Black Holes}.
\newblock {\em The Astrophysical Journal}, 739(2):86, October 2011.

\bibitem{moore+15}
C.~J. {Moore}, R.~H. {Cole}, and C.~P.~L. {Berry}.
\newblock {Gravitational-wave sensitivity curves}.
\newblock {\em Classical and Quantum Gravity}, 32(1):015014, January 2015.

\bibitem{robson+19}
Travis {Robson}, Neil~J. {Cornish}, and Chang {Liu}.
\newblock {The construction and use of LISA sensitivity curves}.
\newblock {\em Classical and Quantum Gravity}, 36(10):105011, May 2019.

\end{thebibliography}

\newpage

\newpage

\end{document}